\documentstyle[prb,preprint,aps,eqsecnum,epsf]{revtex}

\begin{document}

%=====================================================================
\title{Electronic Ladders with SO(5) Symmetry: Phase Diagrams and
       Correlations at half-filling}
%%
%% Author(s):
%%
\vspace{1.5 em}

\author{Holger Frahm and Martin Stahlsmeier}
\address{Institut f\"ur Theoretische Physik, Universit\"at Hannover,
	 D-30167~Hannover, Germany}
\date{\today}
\maketitle
\begin{abstract}
We construct a family of electronic ladder models with $SO(5)$ symmetry
which have exact ground states in the form of finitely correlated wave
functions.  Extensions for these models preserving this symmetry are
studied using these states in a variational approach.  Within this
approach, the zero temperature phase diagram of these electronic
ladders at half filling is obtained, reproducing the known results in
the weak coupling (band insulator) and strong coupling regime, first
studied by Scalapino, Zhang and Hanke.  Finally, the compact form of
the variational wave functions allows to compute various correlation
functions for these systems.
\end{abstract}
%\pacs{
%05.30.-d % Quantum statistical mechanics
%71.10.Pm % Fermions in reduced dimensions (anyons, composite fermions,
%         % Luttinger liquid, etc.)
%75.10.Jm % Quantized spin models
%11.25.Hf % Conformal field theory, algebraic structures
%}
%=====================================================================

%\begin{multicols}{2}

%%%%%%%%%%%%%%%%%%%%%%%%%%%%%%%%%%%%%%%%%%%%%%%%%%%%%%%%%%%%%%%%%%%%%%

\section{Introduction}
The use of symmetries is an important tool to understand the effects
of strong correlation in electronic systems.  Recently, the
$SO(3)$-symmetry of the antiferromagnetic (AFM) order parameter has
been combined with that of $d$-wave superconductivity to form a
five-component vector order parameter \cite{Zhang97}.  It has been
argued that the low energy sector of the resulting theory exhibits an
approximate $SO(5)$-symmetry which allows to explain certain features
such as the vicinity AFM order and superconductivity in the phase
diagram of the high-$T_c$ materials.
Numerical diagonalization studies have been performed and the spectrum
of low lying excitations could in fact be classified according to this
symmetry.

A complementary approach has been the attempt to construct microscopic
electronic systems with manifest $SO(5)$ invariance and studies of
such models to extract the low energy behaviour.  Scalapino, Zhang and Hanke
succeeded in constructing a two-chain ladder Hamiltonian of this type
and studied the strong coupling phase diagram of this system where
they were able to identify several distinct phases
(Ref.~\onlinecite{SZH98}, referred to as SZH in the following).  The
properties of these systems at weak coupling in the metallic regime
has been studied by means of bosonization \cite{SheSen98,LiBF98}.
Such ladder systems, in particular for magnetic insulators have
attracted much attention recently due to the existence of various
experimental realizations in materials closely related to the
high-$T_c$ substances \cite{DaRi96}.  An interesting observation of
Ref.~\onlinecite{SZH98} is the existence of an $SO(5)$ superspin phase
which has been studied in a variational approach based on finitely
correlated matrix product states similar to the ones used for $S=1$
Haldane magnets \cite{AKLT87,fnw:89,ksz:91}.  Finitely correlated states 
have also been considered in electronic systems to describe aspects of 
the phase diagram of extended Hubbard models \cite{Na00,NaI00} and other 
one-dimensional electronic models \cite{Dko00:2,Kim99}.  

For $SU(2)$ spin systems the variational approach has been generalized
to lattices with ladder geometry and proven to give access to large
parts of their phase diagram \cite{Komi98:1,BAMN98,NeMi96,ToSu95}.  
This is the motivation for the present work where we extend the matrix 
product states originally introduced in Ref.~\onlinecite{SZH98} to describe 
the strong coupling physics of the $SO(5)$ superspin phase.  We construct
manifestly $SO(5)$ invariant many particle wave functions from
matrices containing \emph{all} 16 electronic states on a given rung of
the electronic ladder.  The relative weight of the six different
$SO(5)$ multiplets on a rung is controlled by free parameters which
are used to perform a variational study of the zero temperature phase
diagram of the ladder at half filling.  At strong coupling the results
known from Ref.~\onlinecite{SZH98} are reproduced within our approach.
Furthermore, at weak coupling and sufficiently large interchain
hopping amplitude $t_\perp$ the matrix product state correctly
describes the gapped ground state of a band insulator corresponding to
a filled Fermi sea of electrons with one parity.  For intermediate
coupling we find a phase with finite amplitude of the $SO(5)$ spinor
quartets which are essential for the presence of a metallic phase of
the ladder.  The compact form of the variational states allows to
study various correlation functions of interest.

In the following section 
%\ref{sec:states} 
we present the classification of the electronic states of a two-leg
ladder system according to the $SO(5)$-symmetry and discuss all
possible $SO(5)$ symmetric single rung interactions. In Section
\ref{sec:szh} we review the SZH-model and consider tensor products of
rung states to include couplings of neighboring rungs.  Section
\ref{sec:ext} deals with various $SO(5)$ symmetric extensions of this
model and a general construction routine for systems with exact finitely
correlated ground states is given.  Section \ref{sec:var} contains a
detailed analysis of the ground state phase diagram of the system in
the case of weak and intermediate coupling within a variational
approach based on such wave functions.  Furthermore we calculate the
corresponding correlation functions within this approach.  A summary
of our results is given in Section \ref{sec:sum}.

%%%%%%%%%%%%%%%%%%%%%%%%%%%%%%%%%%%%%%%%%%%%%%%%%%%%%%%%%%%%%%%%%%%%%%%

\section{Electronic States of SO(5)-Symmetric Ladder Models}
\label{sec:states}
We consider a two-chain electronic ladder model with canonical creation and
annihilation operators $c_{\sigma}^{\dagger} \left( x \right)$, $c_{\sigma}
\left( x \right)$ for electrons (with spin-projection $\sigma =
\uparrow,\downarrow$) on sites $x$ of the upper leg and analogous operators
$d_{\sigma}^{\dagger}\left( x \right) ,d_{\sigma} \left( x \right)$ for the 
electrons on the lower leg .  In order to discuss the $SO(5)$ symmetry of the
ladder model and to classify all the $16$ possible states on a rung according
to this symmetry, these operators are combined into four-dimensional $SO(5)$ 
spinors \cite{SZH98,RaDe98}

\begin{eqnarray}
 \Psi_{\alpha} \left( x \right) = \left( 
    \begin{array}{cccc} 
   c_{\uparrow} \left( x \right), & 
   c_{\downarrow} \left( x \right), & 
   d_{\uparrow}^{\dagger} \left( x \right), &
   d_{\downarrow}^{\dagger} \left( x \right)
    \end{array} \right)^T \quad \mbox{($x$ even)}
\label{spinor1}
\end{eqnarray}
and
\begin{eqnarray}
 \Psi_{\alpha} \left( x \right) = \left( 
    \begin{array}{cccc} 
   d_{\uparrow} \left( x \right), & d_{\downarrow} \left( x \right), &
   c^{\dagger}_{\uparrow} \left( x \right), c^{\dagger}_{\downarrow}
 \left( x \right)  
    \end{array} \right)^T \quad \mbox{($x$ odd).} 
\label{spinor2}
\end{eqnarray}
Using this definition the ten local generators $L_{ab}$ of the $SO(5)$-algebra
on a single rung $x$ are defined as
\begin{equation}
 L_{ab}(x)=-\frac{1}{2} \Psi_{\alpha}^{\dagger} \left(x\right) 
            \Gamma^{ab}_{\alpha\beta} \Psi_{\beta}\left(x\right), 
       \quad a,b = 1,\ldots,5\ .
\end{equation}
Here $\Gamma^{ab}$ are ten antisymmetric, $4\times4$ matrices (their explicit
form is given in Appendix \ref{app:gam}).  A convenient basis of the Hilbert
space on a single rung is diagonal in the quadratic Casimir charge
\begin{equation}
   C(x)= \sum_{a<b} L^2_{ab}(x).
\label{casim2}
\end{equation}  
In addition we choose to diagonalize the total charge
$Q=\frac{1}{2}(c^{\dagger}c+ d^{\dagger}d-2)$ and the z-component of the spin
$S^z=\frac{1}{2}(c^{\dagger}\sigma_zc+ d^{\dagger}\sigma_zd)$.  Based on the
eigenvalues of $C$ the Hilbert space can be decomposed into six $SO(5)$
multiplets:
\begin{itemize}
\item  Three $SO(5)$ singlets ($C=0$), for $R$ see (\ref{Rmat})
\begin{eqnarray}
 |\Psi^{(1)}_{0,0} \rangle &=& \left| \Omega \right\rangle \; \equiv \; 
 \frac{c^{\dagger}_{\uparrow}d^{\dagger}_{\downarrow}
      -c^{\dagger}_{\downarrow}d^{\dagger}_{\uparrow}}{\sqrt{2}} 
     \left| 0 \right\rangle \; = \; \frac{1}{\sqrt{2}}  
     \left( \left| { \uparrow \atop \downarrow } \right\rangle - \left| 
     { \downarrow \atop \uparrow } \right\rangle \right) 
\nonumber\\
 |\Psi^{(2)}_{0,0} \rangle &=&
 \frac{1}{\sqrt{8}} \Psi_{\alpha} R_{\alpha \beta} \Psi_{\beta}  
     \left|  \Omega \right\rangle  \sim \left| 
   { \uparrow \downarrow \atop - } \right\rangle 
\label{sing1}\\
 |\Psi^{(3)}_{0,0} \rangle &=&
 \frac{1}{\sqrt{8}} \Psi^{\dagger}_{\alpha} R_{\alpha \beta} 
    \Psi^{\dagger}_{\beta}  
   \left|  \Omega \right\rangle \sim  
	\left| { - \atop \uparrow \downarrow } \right\rangle .
\nonumber
\end{eqnarray}

\item  An $SO(5)$ vector quintet ($C=4$) containing the ferromagnetically
polarized state at half filling
\begin{equation}
 |\Psi^{(1)}_{5,\alpha}\rangle \in
 \left\{ \left| { -  \atop -} \right\rangle,  
 \; \left| { \uparrow \downarrow \atop \uparrow \downarrow } \right\rangle,
 \;  \left| { \uparrow \atop \uparrow } \right\rangle, 
 \; \left| { \downarrow \atop \downarrow } \right\rangle,
 \; \left| { \uparrow \atop \downarrow } \right\rangle
 + \left| { \downarrow \atop \uparrow } \right\rangle
 \right\}\,
\quad \alpha=1,\ldots,5.
\label{quin1}
\end{equation}

\item Two $SO(5)$ spinor quartets ($C=5/2$) for an odd number of electrons on
a given rung
\begin{eqnarray}
 |\Psi^{(1)}_{4,\alpha} \rangle &\sim&
 \sqrt{2}\; \Psi_{\alpha} \left|  \Omega \right\rangle \in
 \left\{  \left| { - \atop \uparrow } \right\rangle, 
 \; \left| { - \atop \downarrow } \right\rangle , 
 \; \left| { \uparrow  \atop \uparrow \downarrow } \right\rangle , 
 \; \left| { \downarrow  \atop \uparrow \downarrow } \right\rangle
 \right\}, \quad \alpha=1,\ldots4,
\nonumber\\
\label{quar1}\\
 |\Psi^{(2)}_{4,\alpha} \rangle &\sim&
 \sqrt{2} \; \Psi^{\dagger}_{\alpha} \left|\Omega 
\right\rangle   \in
\left\{  \left| { \uparrow \atop - } \right\rangle, 
\; \left| { \downarrow \atop - } \right\rangle , 
\; \left| { \uparrow \downarrow   \atop \uparrow  } \right\rangle , 
\; \left| { \uparrow \downarrow  \atop \downarrow } \right\rangle
 \right\}, \quad \alpha=1,\ldots4.
\nonumber
\end{eqnarray}

\end{itemize}
We label the states $|\Psi^{(k)}_{d,\alpha} \rangle$ on a rung by the
dimension $d$ of the corresponding multiplet ($\alpha=1,\ldots,d$) and
an additional index $k$.  Similarly, we can characterize product states 
on two rungs (see Sect. \ref{sec:szh}).  Alternatively, the vector
quintet (\ref{quin1}) can be constructed from $SO(5)$ spinors
$|\Psi_{5,1(2)}^{(1)}\rangle = \frac{1}{\sqrt{2}} (n_1\pm n_5)
|\Omega\rangle$, $|\Psi_{5,3(4)}^{(1)} \rangle = \frac{1}{\sqrt{2}}
(n_2\pm n_3) |\Omega\rangle$ and $|\Psi_{5,5}^{(1)} \rangle = n_4
|\Omega\rangle$, with the superspin vector 
\begin{equation}
  n^a \left( x \right )
  \equiv \frac{1}{2}\Psi^{\dagger}_{\alpha}\left( x \right)
  \Gamma^a_{\alpha \beta} \Psi_{\beta} \left( x \right),\quad
  a=1,\ldots,5.
\label{so5n}
\end{equation}
Again, the explicit form of the $4\!\times\!4$ Dirac
$\Gamma$-matrices $\Gamma^a$ is given in Appendix
\ref{app:gam}.

Any electronic ladder model with a local $SO(5)$-symmetry on a rung
has to preserve the degeneracy of the energy within the states of  
each single multiplet.  The invariant Hamiltonian on a single rung can 
therefore be written as a sum over projection operators on these states:
\begin{eqnarray} 
 h_x= \lambda_5 \sum_{\mu=1}^5 | \Psi_{5,\mu}^{(1)} \rangle 
\langle\Psi_{5,\mu}^{(1)}  |
    + \sum_{k,l=1}^2 \lambda_4^{(k,l)} \sum_{\mu=1}^4
 | \Psi_{4,\mu}^{(k)} \rangle \langle \Psi_{4,\mu}^{(l)}| 
    + \sum_{k,l=1}^3 \lambda_0^{(k,l)} | \Psi_{0,0}^{(k)} 
\rangle \langle \Psi_{0,0}^{(l)}|,
\label{prorung}
\end{eqnarray}
where $\lambda^{(k,l)}_d=(\lambda^{(l,k)}_d)^*$ because of the
hermiticity of $h_x$.  All $SO(5)$-symmetric terms on a rung can be
expressed using linear combinations of these projection operators,
e.g., the projection operator on the first singlet $|
\Psi^{(1)}_{0,0}\rangle$ is
\begin{eqnarray}
\hat{P}_{0,0}^{1,1}&=& | \Psi_{0,0}^{(1)}\rangle \langle \Psi_{0,0}^{(1)}|= 
-\frac{1}{3}\vec{S_c}(x)\vec{S_d}(x) 
+\frac{4}{3} (\vec{S_c}(x)\vec{S_d}(x))^2, 
\label{pro-sing1}
\end{eqnarray}
with $ \hat{P}_{d,\mu}^{k,l}= | \psi_{d,\mu}^{(k)} \rangle \langle
\psi_{d,\mu}^{(l)} |$ and $\vec{S_c}(x)=\frac{1}{2}c^{\dagger}(x)
\vec{\sigma} c(x)$.  A complete classification of these terms is given
in Appendix \ref{app:ops}.  As a simple example we choose
\begin{eqnarray} \nonumber
\lambda_0= \left( \begin{array}{ccc}  
      \textstyle{-\frac{7}{2}U-3V}  & \textstyle{2 \sqrt{2} t_{\perp}} 
   & \textstyle{-2 \sqrt{2} t_{\perp}} \\ 
   & \textstyle{\frac{U}{2}-V} & \textstyle{   0} \\
                 \ast &     &  \textstyle{\frac{U}{2}-V}    
                 \end{array} \right), \hspace{5mm}
\lambda_4=\left( \begin{array}{cc}  
   \textstyle{0}  & -\textstyle{2 t_{\perp}} \\ \ast &\textstyle{0} 
                 \end{array} \right) 
\hspace{5mm} \mbox{and}\hspace{5mm} \lambda_5=\frac{U}{2}+V, 
\end{eqnarray}
which leads to the Hubbard-type Hamiltonian\cite{SZH98} with an
$SO(5)$-symmetry introduced by SZH \newpage
\begin{eqnarray} \nonumber
 H_{rung}
 &=& H_{Coulomb}+H_{Hopping} \\ \nonumber
 &=& \sum_{x} \left[ 
  \; U \; \left( ( n_{c\uparrow}(x)-\frac{1}{2} )
       ( n_{c\downarrow}(x)-\frac{1}{2} ) + ( c \rightarrow d )\right)
   \right. 
\\ \label{SZH-ham}
 && \hspace{0.8cm} 
      + \;V \; (n_c(x)-1)(n_d(x)-1) 
      +\; J \; \vec{S_c}(x) \vec{S_d}(x)  
\\ \nonumber
 && \hspace{0.8cm} -2t_{\perp}  \left. \left( c_{\sigma}^{\dagger}(x)d_{\sigma}(x)+H.c. \right) \right]  
\label{SZH-hamrung}
\end{eqnarray}
where $J=4(U\!+\!V)$. This condition on the exchange amplitude
guarantees the degeneracy between the states in the $SO(5)$-quintet
and therefore the local $SO(5)$-symmetry of the system.  We will
discuss this model and $SO(5)$ symmetric extensions in the following
sections.
% \ref{sec:szh},\ref{sec:ext},\ref{sec:var}

%%%%%%%%%%%%%%%%%%%%%%%%%%%%%%%%%%%%%%%%%%%%%%%%%%%%%%%%%%%%%%%%%%%%%%
%
\section{Coupling of neighboring rungs}
\label{sec:szh}

\begin{figure}
\epsfxsize=0.4\textwidth
(a) \epsfbox{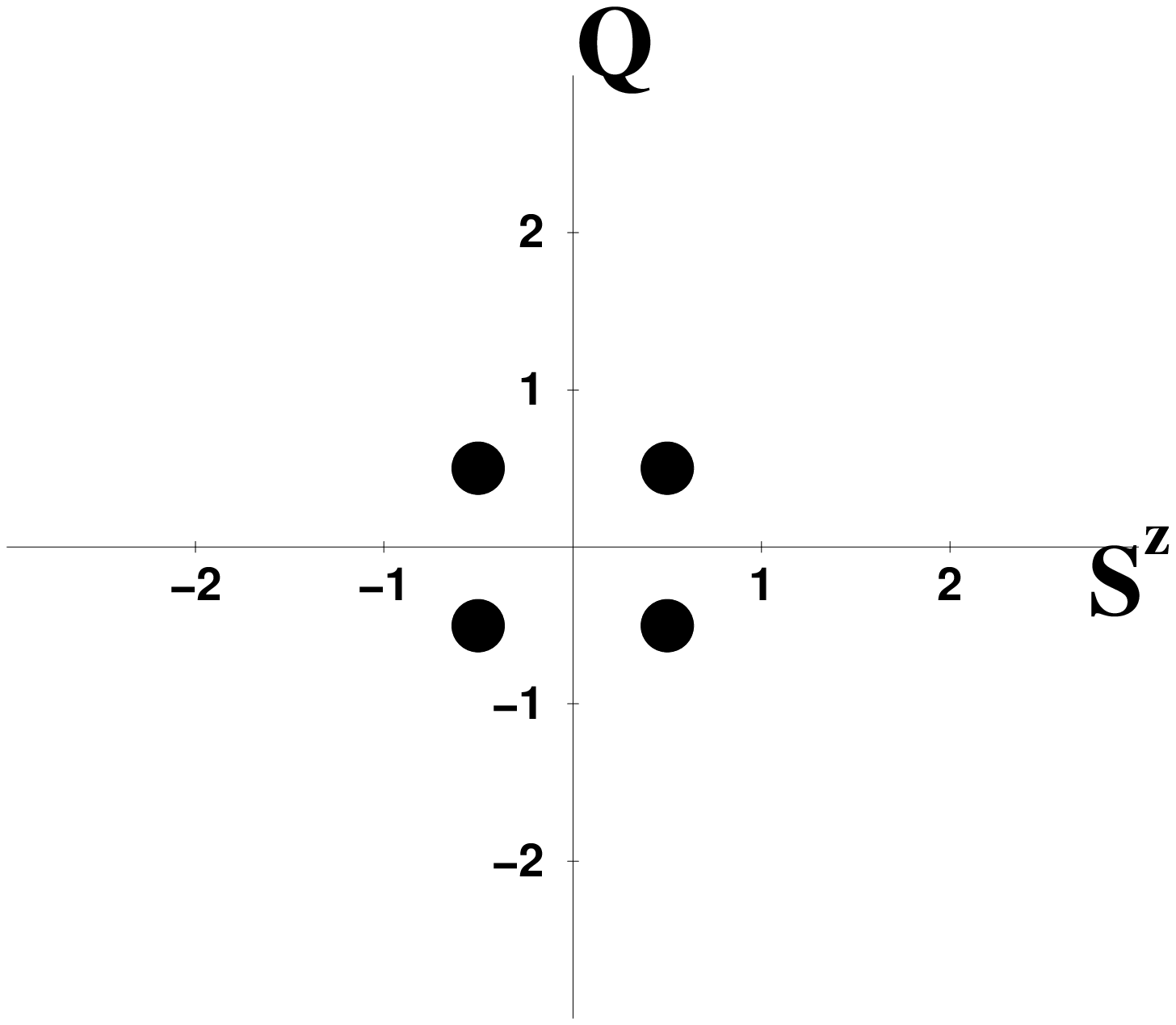} %quartet.eps
\hfill
\epsfxsize=0.4\textwidth
(b) \epsfbox{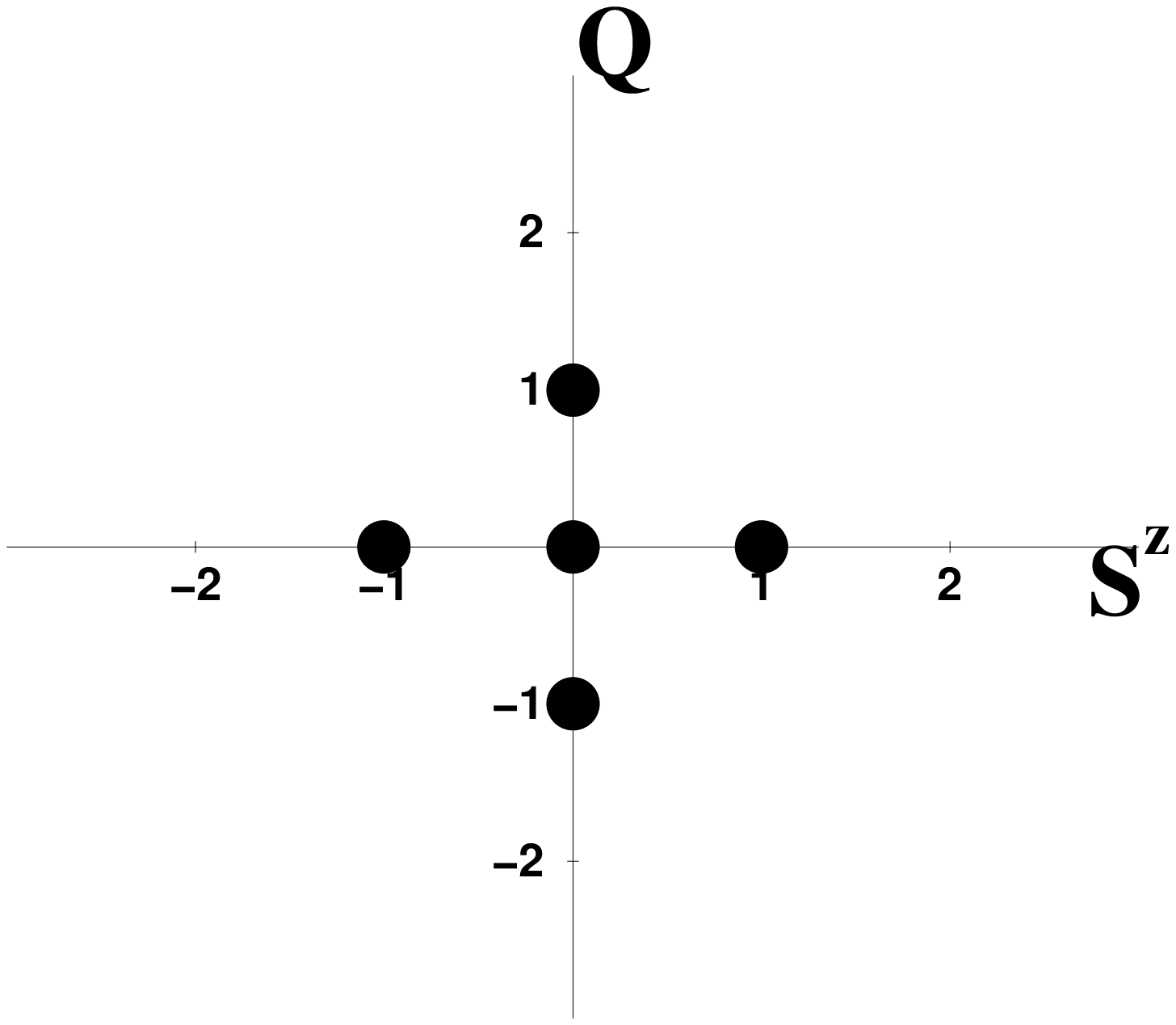} %quintet.eps
\newline
\epsfxsize=0.4\textwidth
(c) \epsfbox{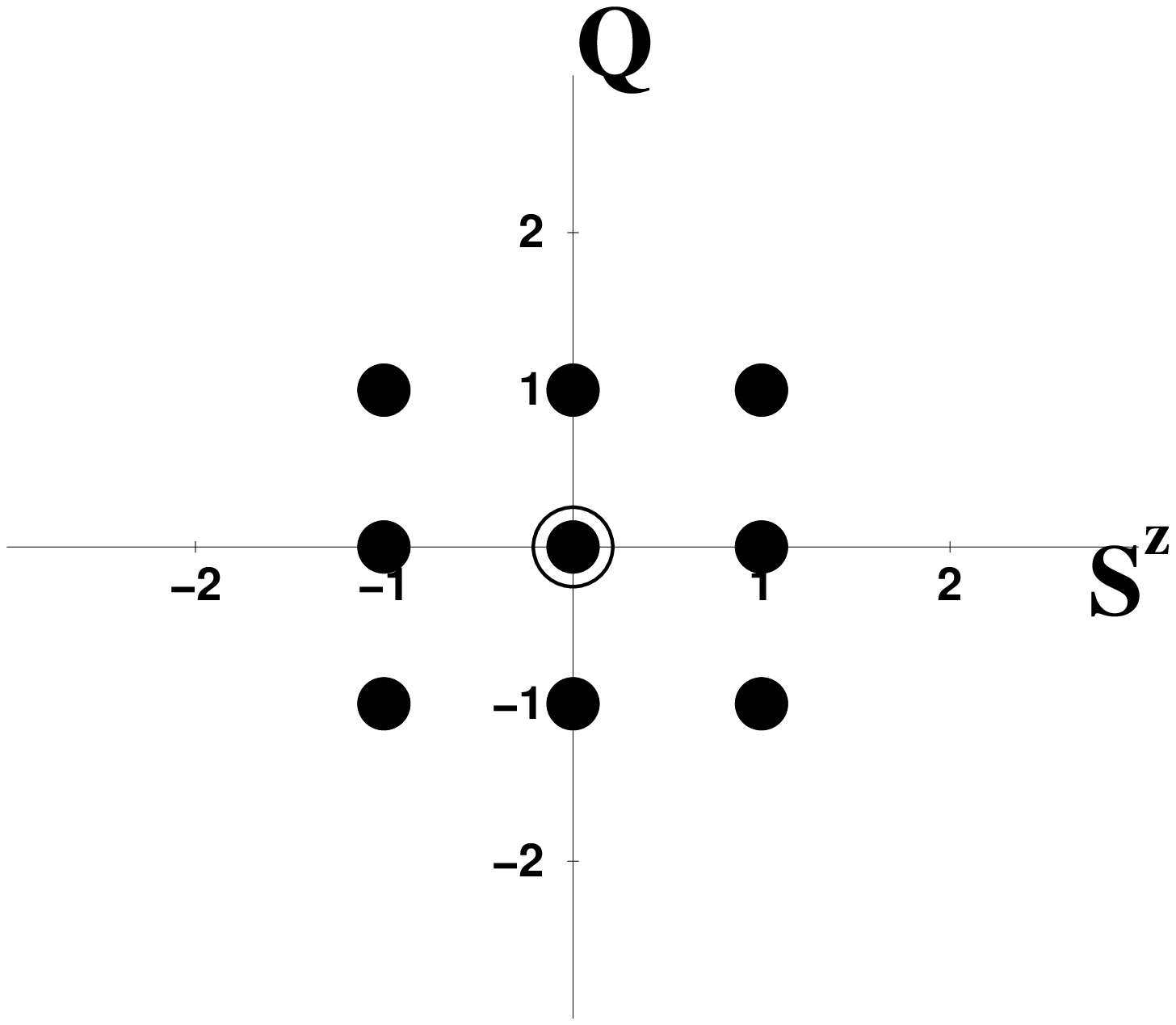} %d10.eps
\hfill
\epsfxsize=0.4\textwidth
(d) \epsfbox{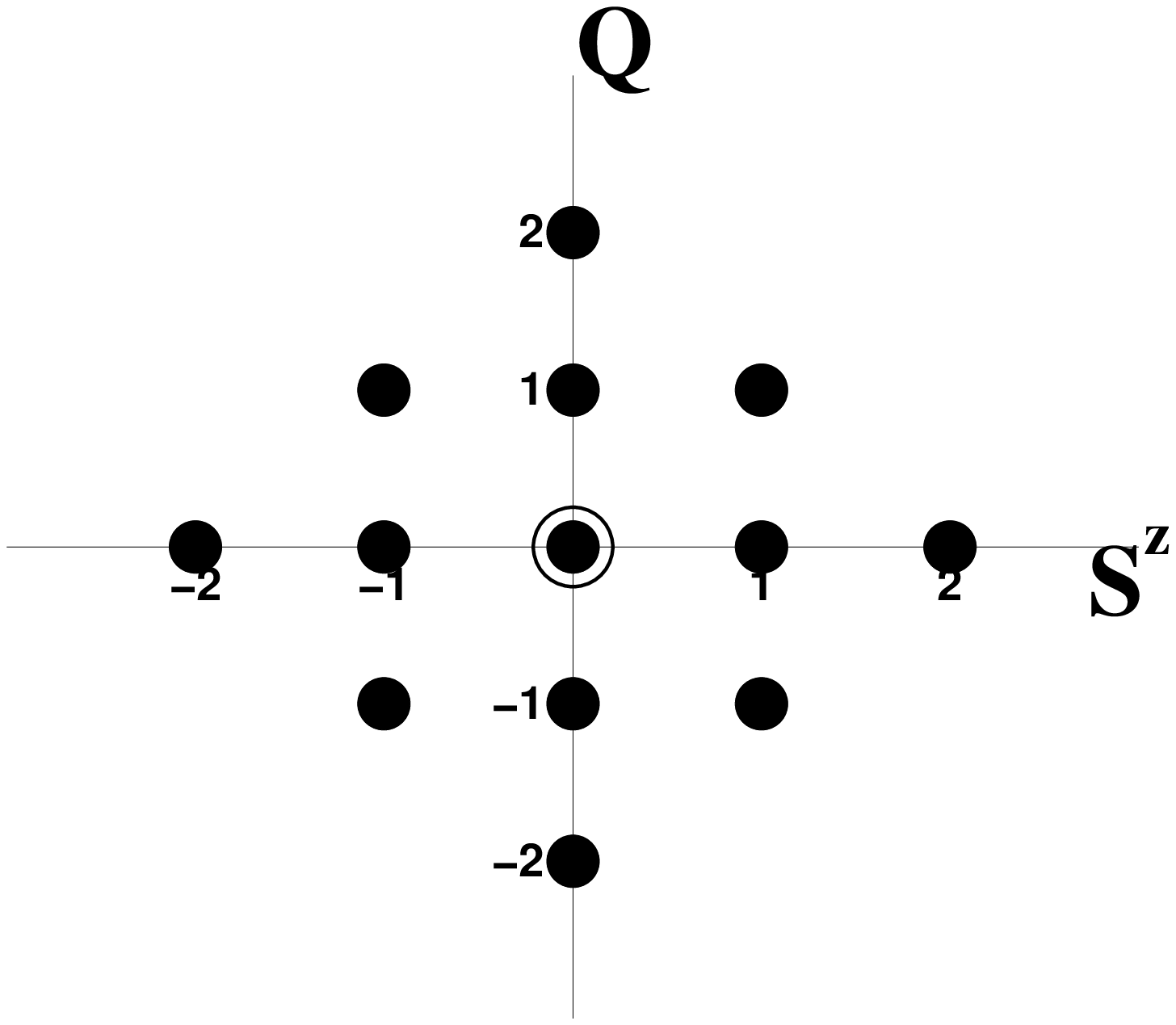} %d14.eps
\newline
\epsfxsize=0.4\textwidth
(e) \epsfbox{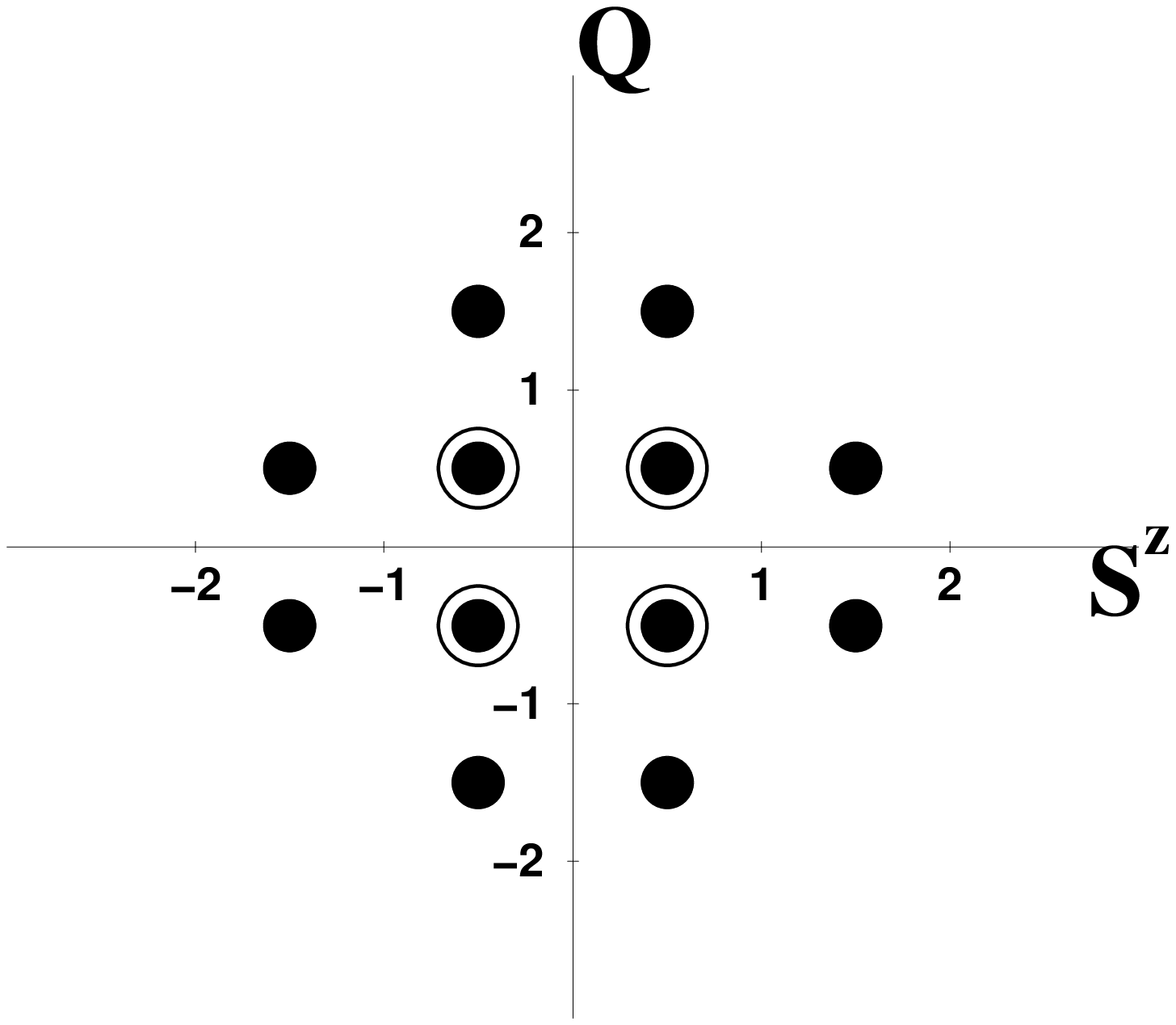} %d16.eps
\caption{The irreducible $SO(5)$ representations appearing on a pair
 of rungs decomposed corresponding to the eigenvalues of $Q$ and
 $S^z$:
 (a) the quartet (with Casimir charge $C=5/2$), (b) the quintet
 ($C=4$), (c) the ten-dimensional ($C=6$), (d) the
 14-dimensional ($C=10$) and (e) 16-dimensional ($C=15/2$) irrep 
 (double circle indicate two states with identical eigenvalues).}
\label{fig1}
\end{figure}

\begin{figure}
\begin{center}
\leavevmode
\epsfxsize=0.7\textwidth
\epsfbox{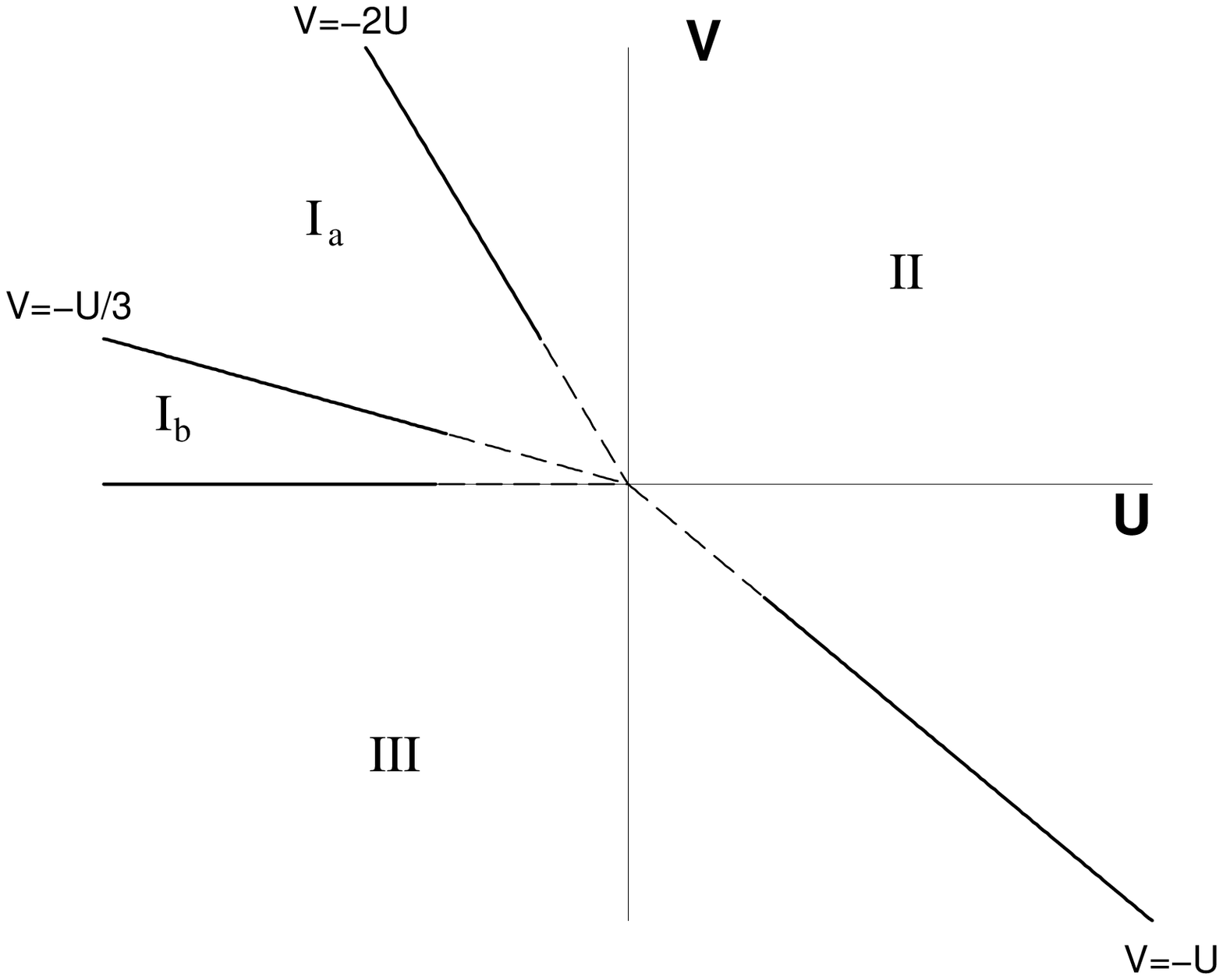}  %strong.eps
\end{center}
\caption{Strong coupling phase diagram, $U$ and $V$ measured in
	 units of $t_{\parallel}$}
\label{fig2}
\end{figure}

In order to describe an extended quasi-one dimensional electronic
system one has to include coupling of neighboring rungs in addition
to single rung interactions considered in the previous section.  The
simplest possible term is an $SO(5)$ symmetric hopping term between
adjacent rungs
\begin{eqnarray}
 -2t_{\parallel} \sum_{\left< x,y \right>}  
\left[ c_{\sigma}^{\dagger}(x)c_{\sigma}(y)
       +d_{\sigma}^{\dagger}(x)d_{\sigma}(y) + H.c. \right],
\label{H-tpar}
\end{eqnarray} 
which can be brought into a manifestly $SO(5)$-symmetric form using
the alternating definitions of the spinors
(\ref{spinor1},\ref{spinor2}).  This hopping term together with the local
rung interactions (\ref{SZH-ham}) yield the complete SZH-model
\cite{SZH98}.  The ground state phase diagram of this system in the
limit of strong coupling ($U,V \gg t_{\perp},t_{\parallel}$) has been
determined by SZH using perturbation theory (see Fig.~\ref{fig2}).
Four different phases have been established at half-filling: \\
In phase $I$ (occuring for $0\le V\le-2U$) the model can be mapped
onto an Ising-like system in a magnetic field: phase $I_a$
($V\ge-U/3$) is a CDW phase and $I_b$ ($V\le-U/3$) corresponds to the
disordered Ising phase.  Phase $II$ is a spin-gap $d$-wave phase
(product of rung singlets), emerging for $V\ge-U,U\ge0$ and for
$V\ge-2U,U\le0$.  The phase $III$ ($V\le-U,V\le0$) is the
superspin-phase where the $SO(5)$-quintet is dominant.  For a further
examination of this superspin phase, SZH have used the finitely
correlated wave function
\begin{equation}
 \left| \Psi_0^{SZH} \right\rangle  = \mbox{Tr} \left( \prod_{x=1}^{L}
	\Gamma^a n_a \left| \Omega  \right\rangle \right)\ 
\label{SZHans}
\end{equation}
(summation over the index $a$ is implied and the trace is taken in the 
$4 \times 4$ matrix space where the $\Gamma^a$ are defined).  In this form 
periodic boundary conditions have been imposed.  By adding many particle
interactions to their original Hubbard-type Hamiltonian this state
(\ref{SZHans}) can be made to be the exact ground state of the 
resulting model. This state has been argued to capture the essential 
physics of the superspin phase --- similar to the r\^ole of the 
AKLT-model as a representative for a Haldane-gapped spin-1 chain.
The wave function (\ref{SZHans}) will be the starting point for 
constructing a generalized matrix product wave function including 
all 16 states on a rung (see section \ref{sec:ext}) and later be used 
for a variational study of the ground state phase diagram of the 
SZH model and its various $SO(5)$ symmetric extensions beyond strong 
coupling (see section \ref{sec:var}).  The hopping term (\ref{H-tpar}) 
is one of many possibilities to include interactions between two adjacent 
rungs of the ladder but the requirement for a local $SO(5)$-symmetry 
puts constraints on the explicit form of these terms.  Explicit expressions
for some of the interaction terms are listed in terms of electron operators
in Appendix \ref{app:ops}.  For a classification of these additional 
interactions we consider products of wave functions on two neighboring 
rungs $x$ and $y$.  A decomposition into $SO(5)$-multiplets similar to 
(\ref{sing1} -- \ref{quar1}) gives 50 different multiplets invariant 
under the action of the $SO(5)$ generators ${\cal L}_{ab}(x,y)=L_{ab}(x)
+L_{ab}(y)$. 
Tensor products containing a singlet factor on one of the rungs are
trivial leading to simple product states, e.g.\ the $SO(5)$ singlets
$|\Psi^{(i)}_{0,0} \rangle_x |\Psi^{(j)}_{0,0} \rangle_y$.
Altogether there are nine singlets, 12 quartets and six quintets of
this form.  The remaining 169 states are obtained by forming tensor
products of quartets (\ref{quar1}) and quintets (\ref{quin1}).  The
decomposition of these products into irreducible representations of
$SO(5)$ reads
\begin{eqnarray*}
 {\bf 4 \otimes 4}&=&{\bf 1 \oplus 5 \oplus 10 }, \\  
 {\bf 4 \otimes 5}&=&{\bf 4 \oplus 16  }, \\
 {\bf 5 \otimes 5}&=&{\bf 1 \oplus 10 \oplus 14 }
\end{eqnarray*}
(numbers denote the dimension of the corresponding $SO(5)$ irrep).
For example, one of four $SO(5)$-singlets in the tensor product of quartet 
states (\ref{quar1}) is
\begin{eqnarray} 
 |\Psi_{0,0}^{(10)}(x,y)\rangle \equiv \frac{1}{2} 
 \left( -|\Psi_{4,1}^{(1)}\rangle |\Psi_{4,3}^{(2)}\rangle 
        -|\Psi_{4,2}^{(1)}\rangle |\Psi_{4,4}^{(2)}\rangle \right.  
 \left. +|\Psi_{4,3}^{(1)}\rangle |\Psi_{4,1}^{(2)}\rangle 
        +|\Psi_{4,4}^{(1)}\rangle |\Psi_{4,2}^{(2)}\rangle \right).
\label{quar2}
\end{eqnarray}
Similar combinations of the rung states appear in the other states,
the Casimir charges of the new multiplets are $C=6$ for the decuplets,
$C=10$ for the 14-dimensional and $C=15/2$ for the 16-dimensional
representations.  The multiplets can be classified further according
to the different eigenvalues of $Q$ and $S^z$ on their member states.
In Fig.~\ref{fig1} the state content of the various multiplets is shown.
In the following we use this classification of the $SO(5)$-multiplets
to construct ladder systems with exact ground states including different
$SO(5)$ symmetric nearest neighbor interactions.

%%%%%%%%%%%%%%%%%%%%%%%%%%%%%%%%%%%%%%%%%%%%%%%%%%%%%%%%%%%%%%%%%%%%%%
%
\section{Extensions of SZH}
\label{sec:ext}
As mentioned in the Introduction the finitely correlated wave
functions originally introduced to discuss the spin-liquid phases
arising in one-dimensional higher spin Heisenberg models
\cite{AKLT87,fnw:89,ksz:91} have recently been generalized to more
general lattices.  In particular, ladder models whose ground states
are of this form have been constructed
\cite{Komi98:1,KoMi97,KoMi98:2}.  In these spin systems the ground
state is of the form $\left| \Psi_0 \right\rangle = \prod_{x=1}^{L}
g_x $ where $g_x$ is a ($2 \times 2$) matrix containing the different
states on a single site or rung $x$ (e.g.\ spin-1 states for the AKLT
model, singlet and triplet states for a two-leg $S=1/2$ ladder, etc.).
Different properties under translation in the extended direction can
be realized by an appropriate choice of the free parameters appearing
in $g_x$ (e.g.\ an alternation to introduce dimerization
\cite{KoMi97,BrMN96,KoMi98:0}).  Within a transfer matrix approach it
is straightforward to compute various ground state correlation
functions for different boundary conditions, periodic ones correspond
to taking the trace of the matrix product wave function \cite{ksz:92}.

For a further analysis of the SZH-model and the construction of
$SO(5)$-symmetric ladder systems with exact ground states in matrix
product form we have extended the wave function (\ref{SZHans}) to
include the three $SO(5)$-singlets (\ref{sing1}) and the two $SO(5)$
spinor quartets (\ref{quar1})
\begin{eqnarray}
 \left| \Psi_0 \right\rangle 
	 = \mbox{Tr} \left( \prod_{x=1}^{L} g_x(\left\{ p_i \right\}) 
	\right).
\label{ansatz}
\end{eqnarray}
Now $g_x$ is a $5\!\times\!5$-matrix and $p_i \; (i=1,...,6)$ are
variational parameters assigning different weights to the multiplets
(\ref{sing1}) -- (\ref{quar1}) on a rung (see Appendix \ref{app:gam}).
We restrict ourselves to the translational invariant case, where the
parameters $p_i$ are chosen to be independent of the rung position
$x$.  In this case the matrix product wave function on two neighboring
rungs contains two $SO(5)$ singlets, two quartets, one quintet and one
decuplet.  The 14-dimensional and 16-dimensional representations are
absent by construction.  The states of the matrix product are linear
combinations of the basis in Sect.~\ref{sec:szh} above, their explicit
form is rather complicated.  With respect to the spin-$SU(2)$
subalgebra the remaining multiplets present in the matrix product
contain spin singlet, doublet and triplet states only (states with
total spin polarization $S^z>1$ are members of the 14 and
16-dimensional representation, see Fig. \ref{fig1}).  An immediate
consequence is that the ansatz cannot be expected to describe the
formation of ferromagnetic domains with higher spin states.  An
analogous argument holds for higher values of the charge $Q$,
corresponding to strong local deviations from half filling.

There is a simple way to construct spin ladder systems with matrix
product wave functions as ground states \cite{Komi98:1} and a
generalization to electronic ladder models with an $SO(5)$-symmetry is
straightforward.  The starting point is a general $SO(5)$-symmetric
Hamilton operator on two neighboring rungs
\begin{eqnarray} \nonumber
     h_{x,x+1}&=& \sum_{k,l=1}^4 \lambda_{16}^{(k,l)}
	 \sum_{\mu=1}^{16} \hat{P}_{16,\mu}^{k,l}
  + \sum_{\mu=1}^{14} \hspace{1mm} \lambda_{14} \, \hat{P}_{14,\mu}^{} 
\\ 
\label{ham-op}
    &+& \sum_{k,l=1}^5 \lambda_{10}^{(k,l)} 
	\sum_{\mu=1}^{10} \hat{P}_{10,\mu}^{k,l} 
  + \sum_{k,l=1}^{10} \lambda_{5}^{(k,l)} \sum_{\mu=1}^{5}
	 \hat{P}_{5,\mu}^{k,l} 
\\ \nonumber
&+& \sum_{k,l=1}^{16} \lambda_{4}^{(k,l)} \sum_{\mu=1}^{4} 
\hat{P}_{4,\mu}^{k,l}
  + \sum_{k,l=1}^{14} \lambda_{0}^{(k,l)} \hat{P}_{0,0}^{k,l}, 
\end{eqnarray}
where $ \hat{P}_{d,\mu}^{k,l}= | \psi_{d,\mu}^{(k)} \rangle \langle
\psi_{d,\mu}^{(l)} |$ are projection operators on all possible
$SO(5)$-multiplets (see section \ref{sec:szh}).  The states
$|\psi_{d,\mu}^{(k)}\rangle$ are product wave functions on two rungs,
$k$ and $l$ label the multiplet, $\mu$ the states in the multiplet and
$d$ is the corresponding dimension of this irreducible
representation. The hermiticity of $h_{x,x+1}$ requires
$\lambda_d^{(k,l)}= (\lambda_d^{(l,k)})^\ast$ for the coupling
constants, leaving altogether 322 free parameters in the Hamiltonian.
A Hamiltonian $H=\sum_x h_{x,x+1}$ has a finitely correlated ground
state $| \Psi_0 \rangle=\prod_x g_x$ with zero energy provided that
the following conditions are satisfied \cite{Komi98:1}:
\begin{itemize}
\item $h_{x,x+1}$ has to annihilate all states contained in the matrix
elements of the product $g_x g_{x+1}$
\item all other eigenstates of $h_{x,x+1}$ have positive energy
\end{itemize} 
Starting with an ansatz for $g_x$ in (\ref{ansatz}) , one has to
identify all multiplets $|\psi_{d,\mu}^{(i_d)}\rangle$ contained in
the product wave function $g_x g_{x+1}$.  These multiplets are
labelled by indices $i_d=1,...,g_d$ where the maximum number $g_d$ is
the number of multiplets with an equal Casimir-charge ($d$ is the
dimension of the irreducible representation), e.g. $g_{10}=2$ if the
product wave function on two neighboring rungs contains two
independent $SO(5)$ decuplets.  After the determination of the
multiplet content of $| \Psi_0 \rangle$ the corresponding parameters
$\lambda_d^{(k,i_d)}$ in $h_{x,x+1}$ are set to zero to fulfill the
first condition. The remaining operators in (\ref{ham-op}) will now
project on states not included in the matrix product wave function,
which leads to zero energy for the ansatz.  To satisfy now the second
condition the reduced matrices $\lambda_d^{(k,l)}$ ($l\ne i_d$) have
to be chosen positive definite (i.e.  positive eigenvalues) such that
(\ref{ansatz}) will be the lowest energy state of the system.

In principle a general Hamiltonian where our ansatz (\ref{ansatz}) is
the exact ground state, can be built by operators projecting on the
other remaining $SO(5)$-multiplets (e.g. the 14-dimensional and the
four 16-dimensional representations) and it has 249 free coupling
constants $\lambda_d^{(k,l)}$ ($l\ne i_d$).  We give explicit
expressions for some of these operators in terms of electron operators
in Appendix \ref{app:ops:gs}.  In general however, the structure of
these projection operators is quite complicated making it difficult to
motivate these exactly solvable systems on physical grounds.

%%%%%%%%%%%%%%%%%%%%%%%%%%%%%%%%%%%%%%%%%%%%%%%%%%%%%%%%%%%%%%%%%%%%%%
%

\section{Variational studies of the phase diagram}
\label{sec:var}

% strong coupling 
An examination of the SZH model beyond strong coupling can be done by using 
(\ref{ansatz}) as a variational wave function. This wave function leads to 
the variational energy
\begin{eqnarray} 
E_{rung}=\langle \Psi_0|H_{Coulomb}|\Psi_0\rangle \sim
(p^2_2+p^2_3)(U/2-V)+5p_6^2(U/2+V)-p^2_1(\frac{7}{2}U+3V)
\label{erung}
\end{eqnarray}
for the spin and charge interaction on a rung (see equation
\ref{SZH-hamrung}).  Here $p_6$ is the parameter corresponding to the
$SO(5)$-quintet, $p_1$ is the weight of the singlet $|\Omega\rangle$
and $p_{2,3}$ of the symmetric and antisymmetric linear combinations
of the other $SO(5)$-singlets (\ref{sing1}).  The variational energy
corresponding to the hopping term on a rung is
\begin{eqnarray}
E_{t_{\perp}}=\langle \Psi_0|H_{Hopping}|\Psi_0\rangle\sim t_{\perp} 
 \left[ 8p_1p_2+2 \left( \frac{p^2_5-p^2_4}{h_2}\right)
 \left( w-h_1\right)\right]
\label{hopp-rung}
\end{eqnarray} 
and between two neighboring rungs (see equation \ref{H-tpar}) it is
\begin{eqnarray}
E_{t_{\parallel}}\sim t_{\parallel} 
  \left[ p_4p_5 \left( 25p_6^2+p_1^2-p_2^2-p_3^2 \right)
 +h_2 \left(p_2p_3-5p_1p_6 \right)
+ \left(p_4^2-p_5^2 \right) \left(5p_2p_6-p_1p_3 \right) \right]. 
\label{hopp-bet}
\end{eqnarray}
where $h_1=5p_6^2+p_1^2+p^2_2+p_3^2$, $h_2=p^2_4+p^2_5$ and
$w=\sqrt{h_1^2+16h_2^2}$.
  
In the strong coupling limit one can neglect these hopping terms
(\ref{hopp-rung}, \ref{hopp-bet}).  Minimizing the energy with respect
to the $p_i$ reproduces exactly the phase diagram calculated by SZH
within perturbation theory (Fig.~\ref{fig2}) where the phases are
fixed by the largest amplitude $p_i$ of the corresponding state and
the crossover is continuous.  The phase $I$ is dominated by the
bonding singlet state with amplitude $p_2$, phase $III$ is the
superspin phase ($p_6$) and phase $II$ consists of products of rung
singlets ($p_1$).  In this approach with translationally invariant
$p_i$ the crossover between the two Ising-phases can not be
reproduced.  We now extend this analysis of the phase diagram of the
SZH model to weak and intermediate coupling.

%%%%%%%%%%%%%%%%%%%%%%%%%%%%%%%%%%%%%%%%%%%%%%%%%%%%%%%%%%%%%%%%%%%%%%
%
\subsection{Weak coupling phase diagram}
\begin{figure}
\epsfxsize=0.4\textwidth
(a) \epsfbox{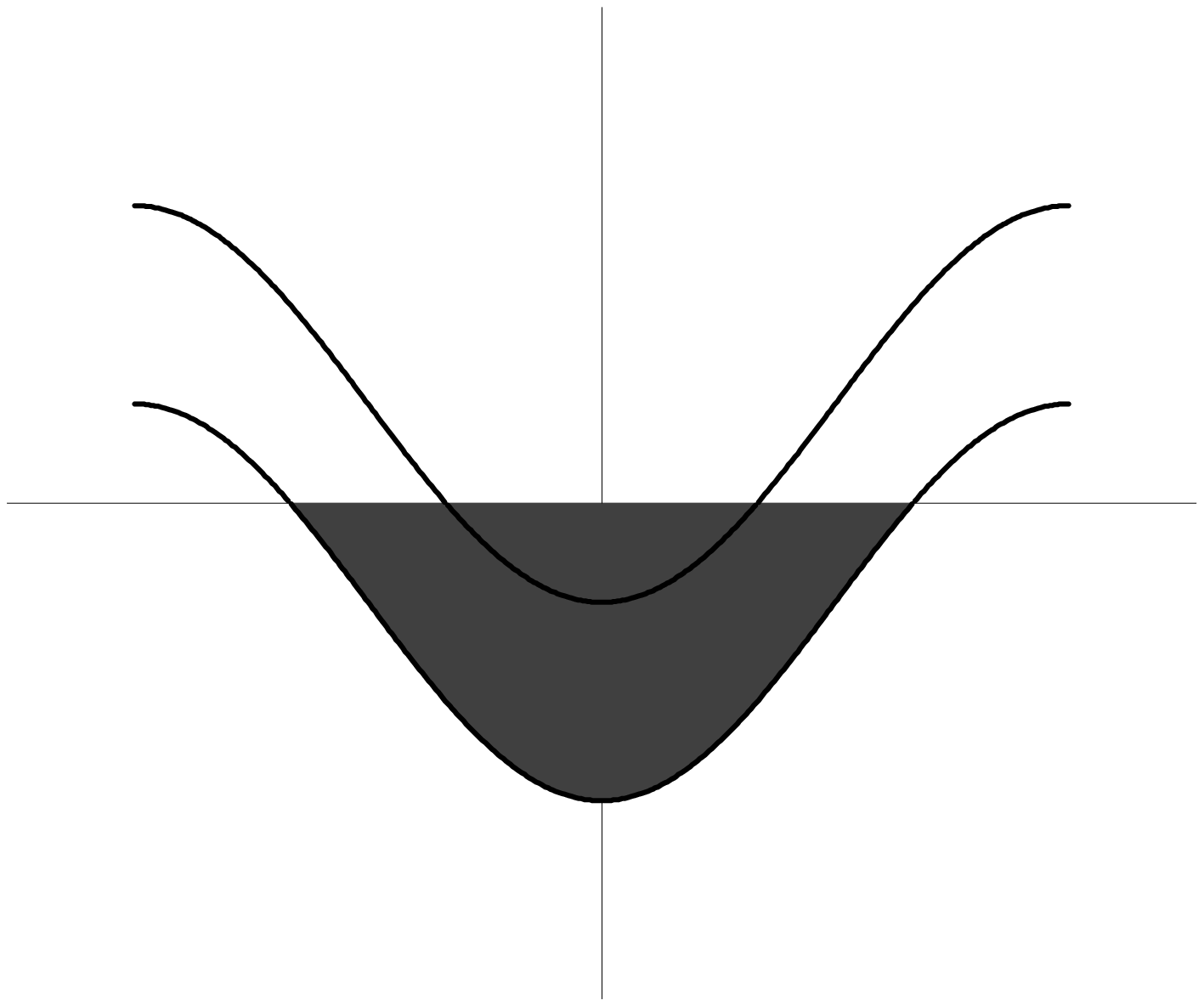}  %11.eps
\hfill
\epsfxsize=0.4\textwidth
(b) \epsfbox{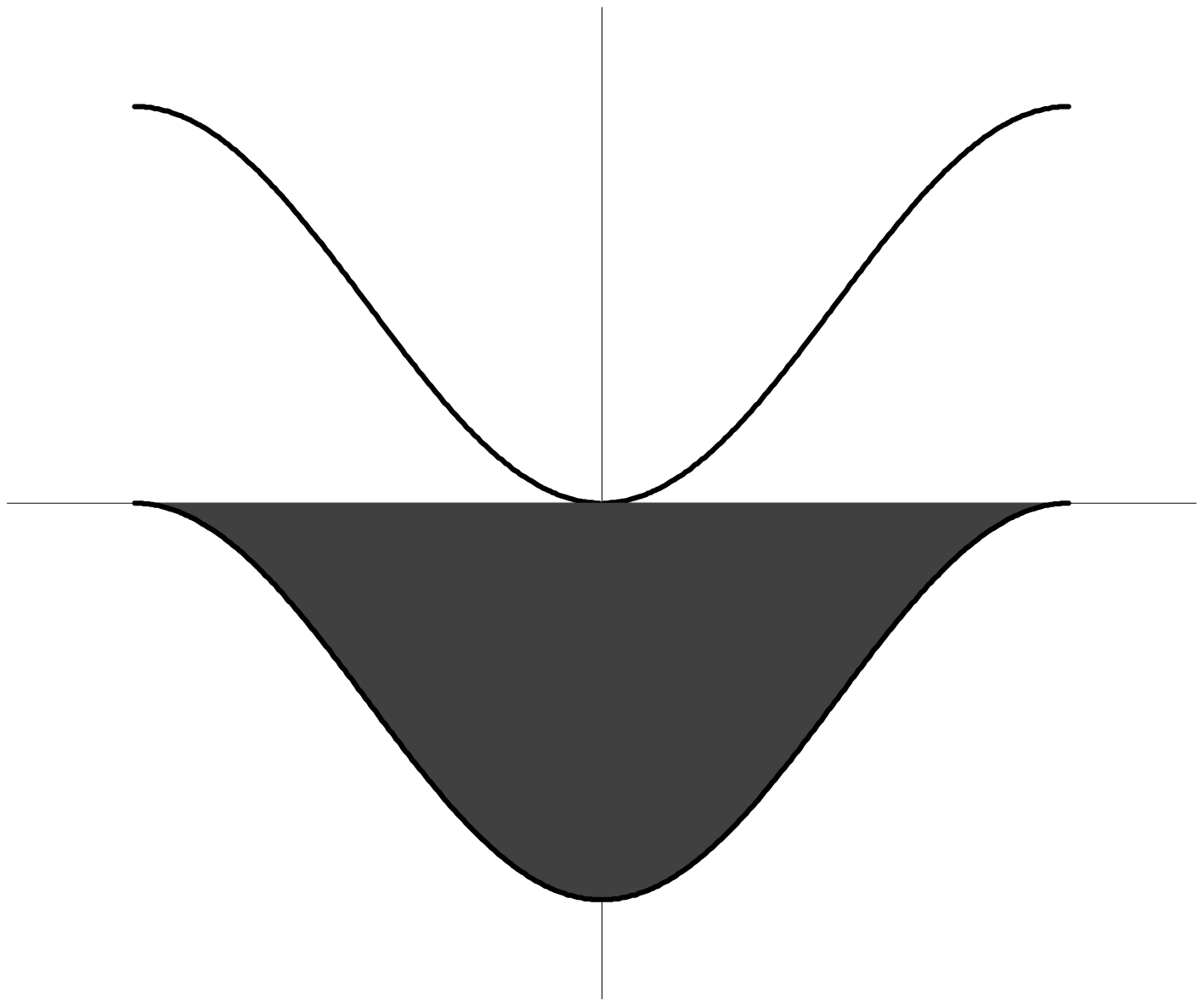} %21.eps
\caption{Band structure of a two-leg ladder model for (a) $t_{\perp} =
t_{\parallel}$ and (b) $t_{\perp} = 2t_{\parallel}$}
\label{fig3}
\end{figure}

The band structure of the non-interacting system at half-filling is
well known.  For $U\!=\!V\!=\!0$ there are two energy bands, given by
\begin{equation}
 \epsilon_\pm(k)=\pm 2t_{\perp}-4 t_{\parallel} \cos(k), 
 \quad -\pi \le k \le \pi
\label{disp0}
\end{equation} 
and two different cases have to be distinguished:

For $t_{\perp}\!<\!2 t_{\parallel} $ the Fermi energy intersects the
two bands (see Fig.~\ref{fig3}.a) and for $t_{\perp}\!\ge\!2
t_{\parallel} $ they are separated by an energy gap (see
Fig.~\ref{fig3}.b).
The gapless system ($t_{\perp}<2 t_{\parallel}$) has been studied
using bosonization of the low lying modes in the vicinity of the four
Fermi points to obtain the phase diagram for weak coupling ($U,V \ll
t_{\perp},t_{\parallel}$): Lin et al.\ \cite{LiBF98} found that at
half filling the system is driven towards an integrable
SO(8)-symmetric Gross-Neveu model in a weak coupling renormalization
group analysis and predicted the occurrence of additional phases
compared to the strong coupling case.  
The ongoing debate of these results (see the criticism of
Ref.~\onlinecite{EMZ99}) cannot be clarified within the present
ansatz: Using finitely correlated wave functions always leads to an
exponential decay of correlation functions, indicating the existence
of an energy gap between the ground state and the first excited state.

%%%%%%%%%%%%%%%%%%%%%%%%%%%%%%%%%%%%%%%%%%%%%%%%%%%%%%%%%%%%%%%%%%%%%%
%

\subsection{Phase diagram for $\bf t_{\perp}\ge2 t_{\parallel}$}
For  $t_{\perp}\ge2 t_{\parallel}$ the variational ansatz gives the
exact ground state for the non-interacting system ($U=V=0$):

Choosing $p_1=-\frac{1}{\sqrt{2}},p_2=\frac{1}{\sqrt{2}}$ and 
$p_i\equiv0$ for $i=3,...,6$ we find
\begin{eqnarray}
\left| \Psi_0 \right\rangle \sim 
\prod_{x=1}^L 
\left(-c^{\dagger}_{\uparrow}d^{\dagger}_{\downarrow}
      +c^{\dagger}_{\downarrow}d^{\dagger}_{\uparrow}
      -d^{\dagger}_{\uparrow}d^{\dagger}_{\downarrow}
      -c^{\dagger}_{\uparrow}c^{\dagger}_{\downarrow}
\right)
 \left( x \right) \left| 0 \right\rangle,
\end{eqnarray}
which corresponds to complete filling of the modes with energy
$\epsilon_-(k)$ in (\ref{disp0}), the band insulator.  Consequently,
we expect the variational approach to give reasonable results for the
weak coupling phase diagram in this regime of hopping amplitudes.  The
quality of the approach can be measured by the mean deviation
$\sqrt{\langle (\Delta H)^2 \rangle} = \sqrt{ \langle H^2 \rangle
-\langle H \rangle^2} $ of the energy.  For $U,V \ll
t_{\perp},t_{\parallel}$ the mean deviation stays small compared to
the energy so that the ansatz should give reliable results.

We find that only two phases are present in the weak coupling case
(see Fig.~\ref{fig4}) : the Ising phase $I$ ($p_2$) and the spin-gap
d-wave phase $II$ ($p_1$) already known from the strong coupling
diagram (see Fig.~\ref{fig2}).  The superspin phase disappears and
also the $SO(5)$-quartets have no significant weight ($p_4,p_5,p_6
\sim 0$) as expected for a band insulator.

Considering the complete ground state phase diagram (see
Fig.~\ref{fig4}) we find an additional phase for intermediate coupling
($U,V \cong t_{\perp},t_{\parallel}$) where the $SO(5)$-quartets have
the largest weight, in particular the rung-symmetric one
$|Q_{\alpha}^+\rangle$ (\ref{eq:quar}).
Apart from these, the symmetric singlet state $( \Psi^{\dagger}_{\alpha}
R_{\alpha \beta} \Psi^{\dagger}_{\beta}-\Psi^{}_{\alpha} R_{\alpha
\beta} \Psi^{}_{\beta}) |\Omega \rangle$ --- which determines the
ground state in phase $I$ --- has a significant weight.
Due to the resonating structure of the ansatz and the relatively large
variational value of $\langle (\Delta H)^2 \rangle$ the phase
boundaries are not very accurate --- for a more detailed study of this
question the present work should be complemented by a numerical
approach.  As discussed earlier it is not possible within this
approach to determine the position of the crossover line between the
two Ising phases, or even whether this transition still occurs for the
case of weak or intermediate coupling.

%%%%%%%%%%%%%%%%%%%%%%%%%%%%%%%%%%%%%%%%%%%%%%%%%%%%%%%%%%%%%%%%%%%%%%
%
\subsection{Ground State Correlations}

The physics in the ground state is determined by ground state
correlation functions, which are easily computed from matrix product
wave functions.  The matrix product ansatz (\ref{ansatz}) with the six
free parameters $p_i$ represents the ground state for a large class of
models.  We have calculated various correlation functions explicitly
in the thermodynamic limit ($L \to \infty$) for this variety of models
(a detailed list is given in Appendix \ref{app:cor}) and we determined
the correlation length and the amplitude of different ground state
correlations for the SZH-model when we used the ansatz as a
variational wave function.

\begin{figure}
\begin{center}
\leavevmode
\epsfxsize=0.7\textwidth
% \epsfbox{Pictures/Try.eps}       
\epsfbox{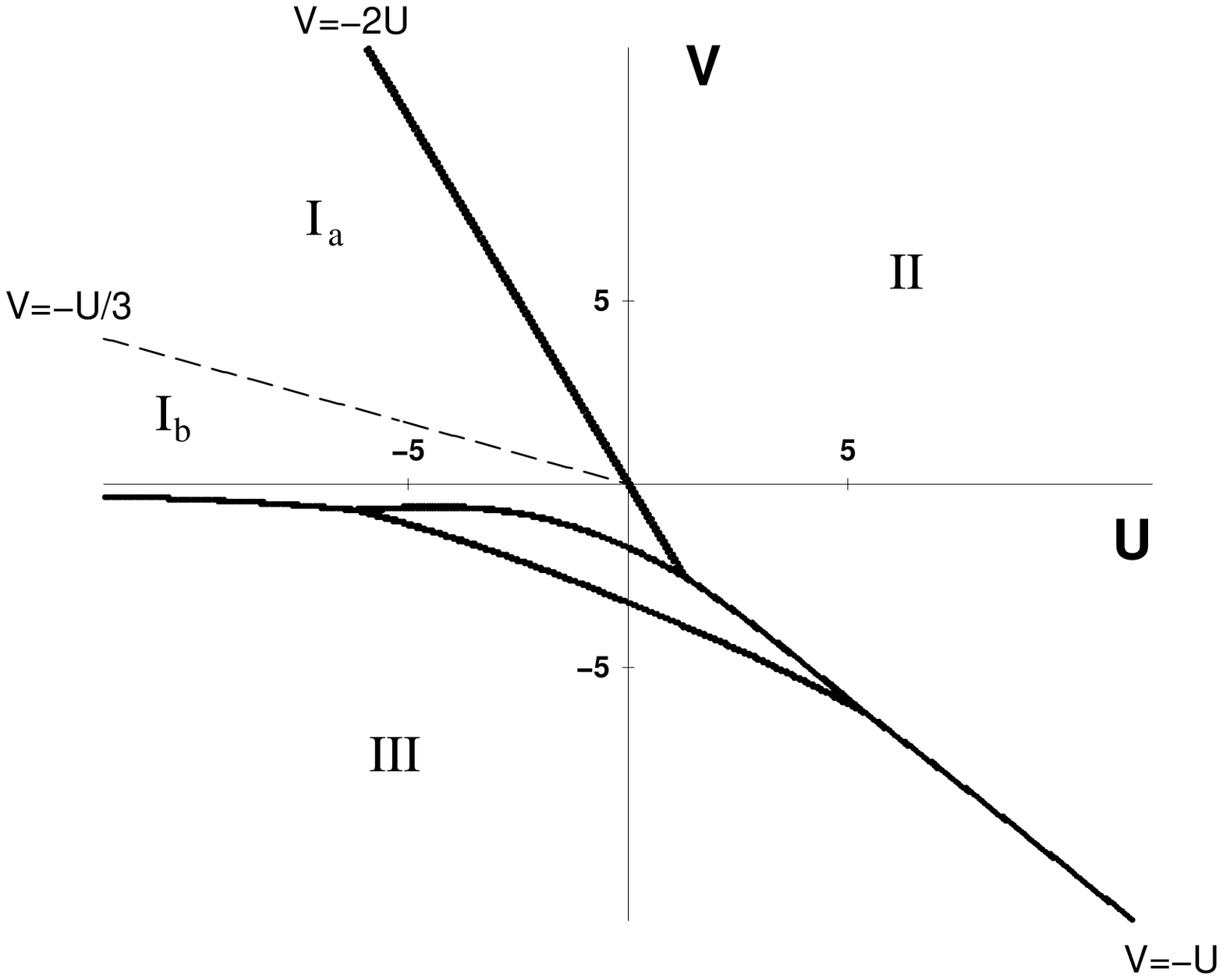} % Try.exact.eps
\end{center}
\caption{Phase diagram for $t_{\perp}=2 t_{\parallel}$ ($t_{\parallel}=1$): the phase 
        boundaries were calculated by comparing the amplitudes of the different multiplets.}
\label{fig4}
\end{figure}

\begin{figure}
\begin{center}
\leavevmode
\epsfxsize=0.7\textwidth
\epsfbox{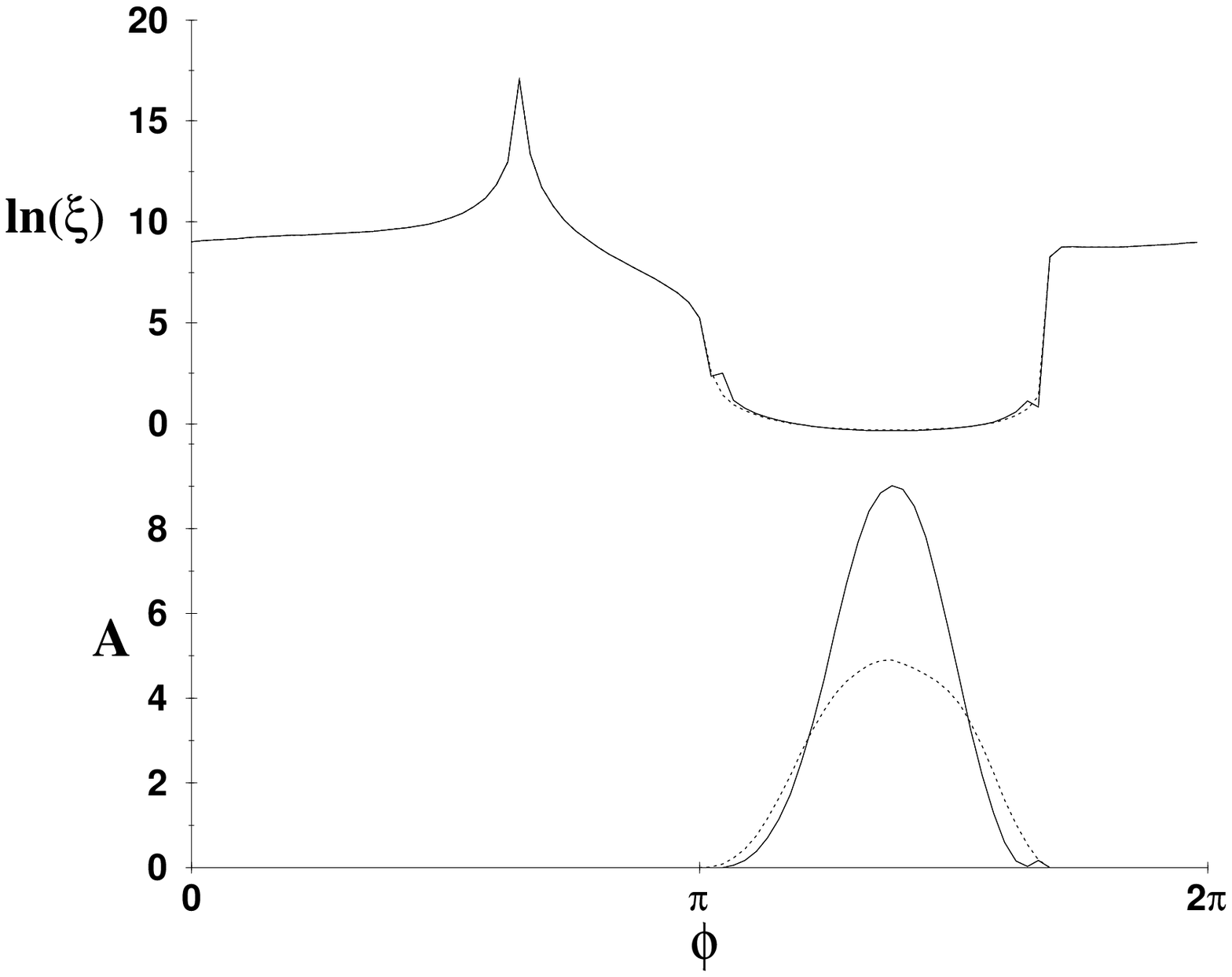} % Spin.eps 
\end{center}
\caption{Correlation length and amplitude for the spin-spin
correlation function , the full line
corresponds to $\langle \vec{S}_{c(d)}(r)\vec{S}_{c(d)}(0) \rangle$
and the dotted to $\langle \vec{S}_{c(d)}(r)\vec{S}_{d(c)}(0)
\rangle$.}
\label{fig5}
\end{figure}

\begin{figure}
\begin{center}
\leavevmode
\epsfxsize=0.7\textwidth
\epsfbox{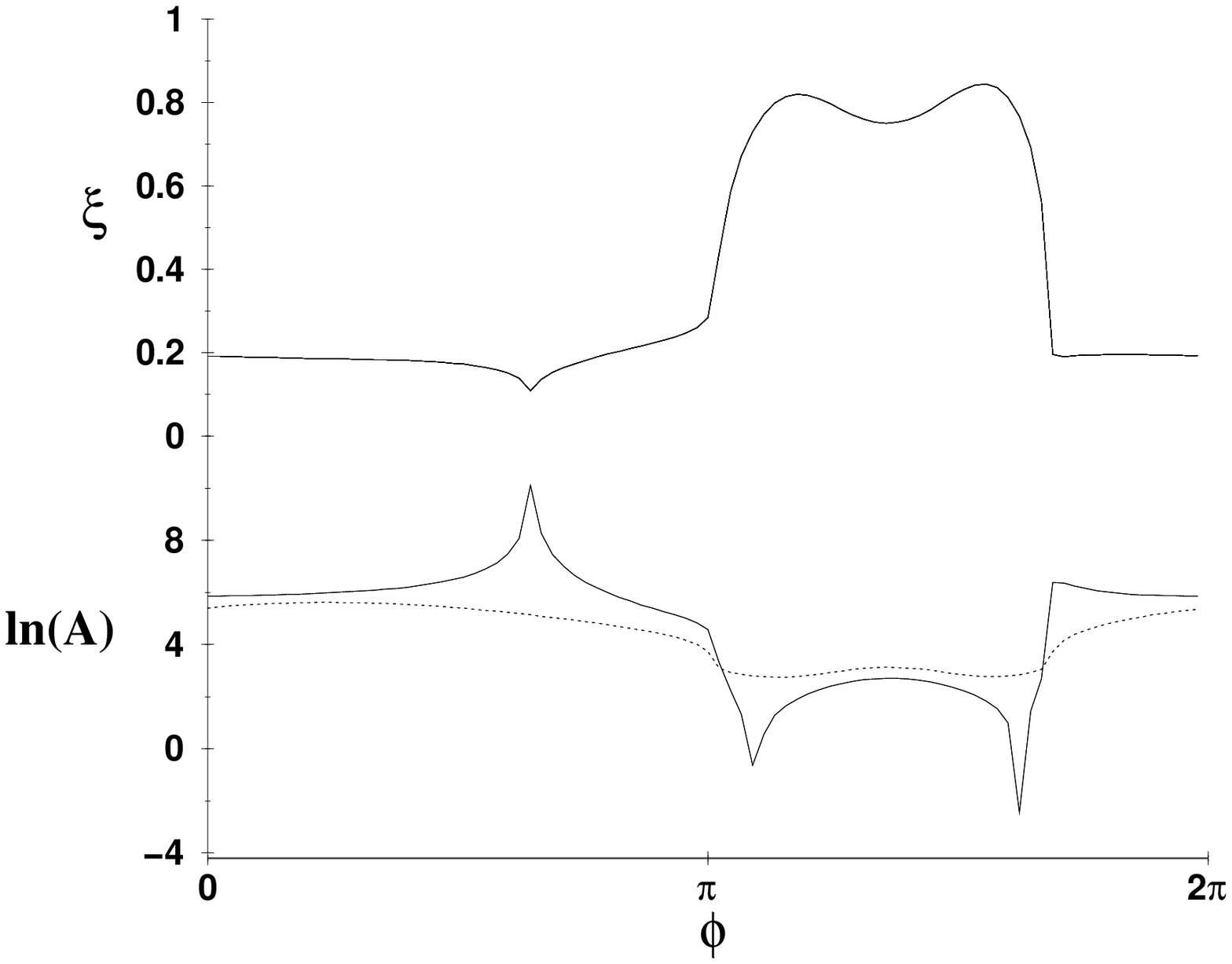} %Elek.eps
\end{center}
\caption{The correlation length and amplitude of the expectation value
$\langle c_{g,u}^{\dagger}(r)c_{g,u}^{}(0) \rangle$, the full line
corresponds to $\langle c_{g(u)}^{\dagger}(r)c_{g(u)}^{}(0) \rangle$
and the dotted to $\langle
c_{g(u)}^{\dagger}(r)c_{u(g)}^{}(0)\rangle$.}
\label{fig6}
\end{figure}

The two-point correlations in matrix product states are always
short-ranged (if not vanishing) and have the following form
\begin{eqnarray*}
\langle O^{\dagger}(r)O(0) \rangle = A(\{p_i\})\; e^{-\frac{r}{\xi}}
\end{eqnarray*}
They exhibit an exponential decay with the correlation length $\xi$
and amplitude $A(\{p_i\})$.  As an example we consider the correlation
length and amplitude of the expectation value of the spin-spin
correlation function $\langle \vec{S}_{c,d}(r)\vec{S}_{c,d}(0)
\rangle$ (see Fig.~\ref{fig5}) and of field correlators $\langle
c_{\alpha}^{\dagger}(r) c_{\beta}^{}(0) \rangle$, $\alpha,\beta \in
\left\{ g,u \right\} $ (Fig.~\ref{fig6}) for the SZH-model on a circle
in the $U$-$V$-plane (with $U^2+V^2=3$) intersecting the phases $I$,
$II$ and the quartet phase (see Fig.~\ref{fig4}).

The spin-spin correlation function $\langle
\vec{S}_{c,d}(r)\vec{S}_{c,d}(0) \rangle$ in Fig.~\ref{fig5} is non-vanishing 
only in the quartet phase but with an extremely small correlation length 
indicating strong nearest neighbor correlations.  For the electron-electron 
correlation in Fig.~\ref{fig6} with 
$c_{g,u}(x)=(c_{\uparrow}(x)+c_{\downarrow}(x)) \pm (d_{\uparrow}(x) +
d_{\downarrow}(x))$ the correlation length $\xi$ is small for all
angles $\phi$ but with a very large amplitude $A$ except in the
quartet phase.

The sharp peak in both diagrams at $\phi\sim\frac{5}{8}\pi$ indicate
the crossover of the phases $II$ and $I$ in Fig.~\ref{fig4} where the
correlation length diverges.  Calculating these correlations in the
strong coupling limit the phase boundaries in Fig.~\ref{fig4} are
denoted by very sharp peaks in the electron-electron correlation
length $\xi_{\langle c_{g,u}^{\dagger}(r)c_{g,u}^{}(0) \rangle}$ with
a non-vanishing amplitude.  The spin-spin correlations are zero in the
whole phase diagram and give no further hints of an underlying
structure in the system.

%%%%%%%%%%%%%%%%%%%%%%%%%%%%%%%%%%%%%%%%%%%%%%%%%%%%%%%%%%%%%%%%%%%
%
\subsection{Variational examination of SO(5)-symmetric extensions}

Our variational approach is also suitable to study the phase diagrams of 
various $SO(5)$-symmetric extensions of the SZH model.  We have considered 
additional interactions on a single rung and between two neighboring rungs,
using the construction routine of section \ref{sec:states} and \ref{sec:ext}.

\subsubsection{Single rung interactions}
\begin{figure}
\epsfxsize=0.4\textwidth
(a) \epsfbox{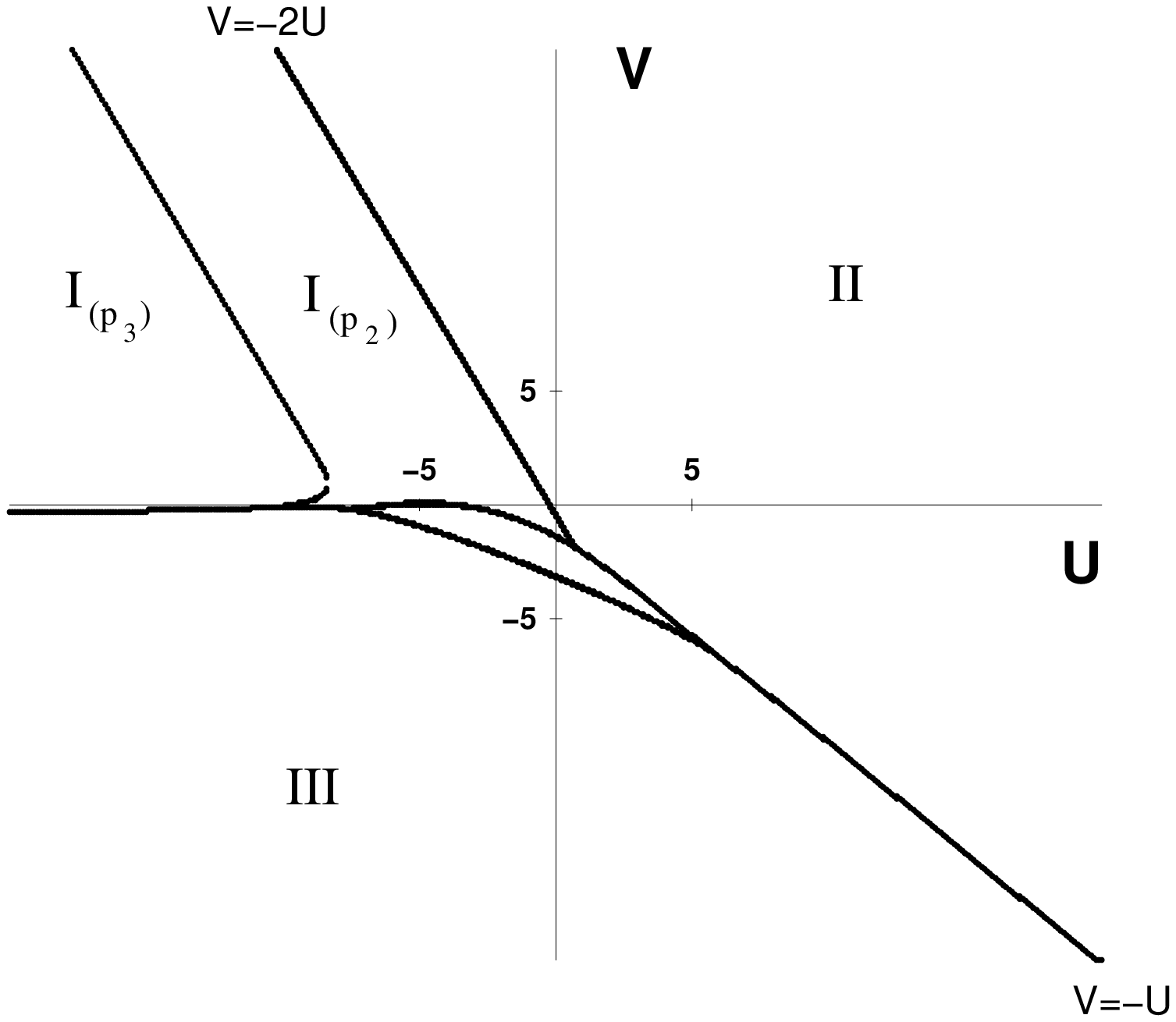} %Paar-1.eps
\hfill
\epsfxsize=0.4\textwidth
(b) \epsfbox{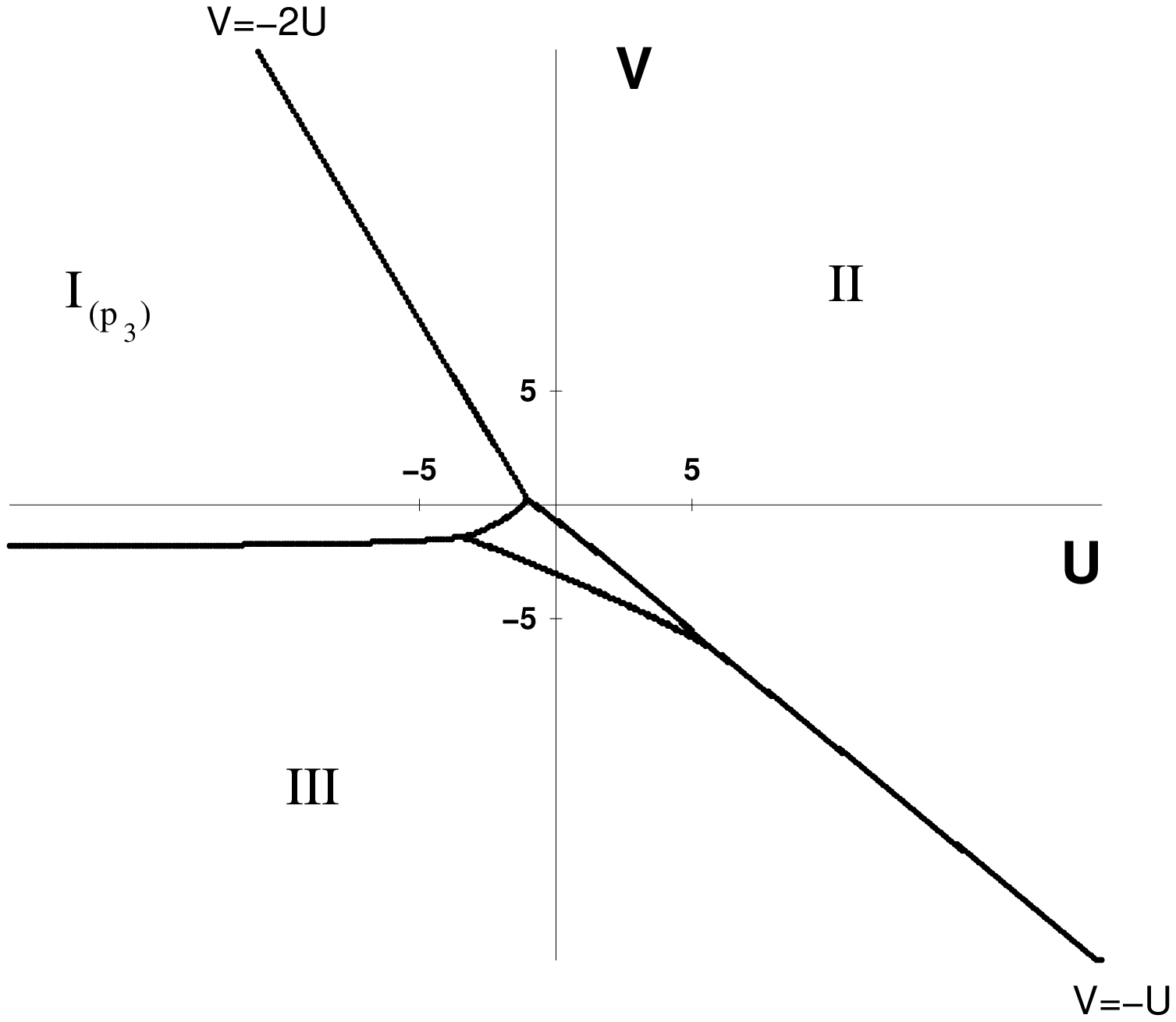} %Paar-4.eps
\caption{Phase diagram for $t_{\perp}=2t_{\parallel}$ including pair hopping 
         for (a) $t_{pair} = -t_{\parallel}$ and (b) $t_{pair} = -4t_{\parallel}$
         with $t_{\parallel}=1$.}
\label{fig7}
\end{figure}

All single rung interactions can be constructed using the projection
operators of Sect. \ref{sec:states} and a detailed list of all
possible terms can be found in the Appendix \ref{app:ops}.  Taking
into account the operators $\hat{P}_{d,\mu}^{k,l}$ with the coupling
constants $\lambda_d^{(k,l)}$ leads to the following contributions to
the variational energy (\ref{erung}), calculated with the ansatz
(\ref{ansatz})
\begin{mathletters}
\begin{eqnarray}
&&\langle \hat{P}^{1,1}_{(0,0)} \rangle = \frac{p_1^2}{w} \\
&&\langle \hat{P}^{2,2}_{(0,0)} \rangle = \frac{1}{2w}(p_2+p_3)^2 \\ 
&&\langle \hat{P}^{3,3}_{(0,0)} \rangle = \frac{1}{2w}(p_2-p_3)^2 \\
&&\langle \hat{P}^{1,2}_{(0,0)}+\hat{P}^{2,1}_{(0,0)} \rangle 
= \frac{\sqrt{2}}{w}p_1(p_3+p_2) \\
&&\langle \hat{P}^{1,3}_{(0,0)}+\hat{P}^{3,1}_{(0,0)} \rangle 
= \frac{\sqrt{2}}{w}p_1(p_3-p_2) \\
&&\langle \hat{P}^{2,3}_{(0,0)}+\hat{P}^{3,2}_{(0,0)} \rangle 
= \frac{1}{w}(p_3^2-p_2^2) \\ 
&&\sum_{\mu} \langle \hat{P}^{1,1}_{(4,\mu)} \rangle 
= \frac{w-h_1}{2wh_2} (p_4-p_5)^2   \\
&&\sum_{\mu} \langle \hat{P}^{2,2}_{(4,\mu)} \rangle 
= \frac{w-h_1}{2wh_2} (p_4+p_5)^2   \\
&&\sum_{\mu} \langle \hat{P}^{1,2}_{(4,\mu)} + \hat{P}^{2,1}_{(4,\mu)}
 \rangle 
= \frac{w-h_1}{wh_2} (p_4^2-p_5^2) \\   
&&\sum_{\mu} \langle \hat{P}^{0,0}_{(5,\mu)} \rangle 
= \frac{5p_6^2}{w}.  
\end{eqnarray} 
\end{mathletters}
These modifications of the model cause some changes in the ground
state phase diagram, e.g. the simple terms like (a) and (j) will only
shift the phase boundaries without changing the general structure of
the phase diagram.  Other interactions like the pair hopping term (f)
\begin{eqnarray}
t_{pair} \left( d_{\uparrow}^{\dagger}d_{\downarrow}^{\dagger}
c_{\uparrow}c_{\downarrow} +h.c. \right)  \sim \hat{P}^{2,3}_{(0,0)}
+\hat{P}^{3,2}_{(0,0)}
\label{ext1}
\end{eqnarray}
will dramatically change the phase diagram (see Fig.~\ref{fig7}).  

\begin{figure}
\epsfxsize=0.4\textwidth
\epsfbox{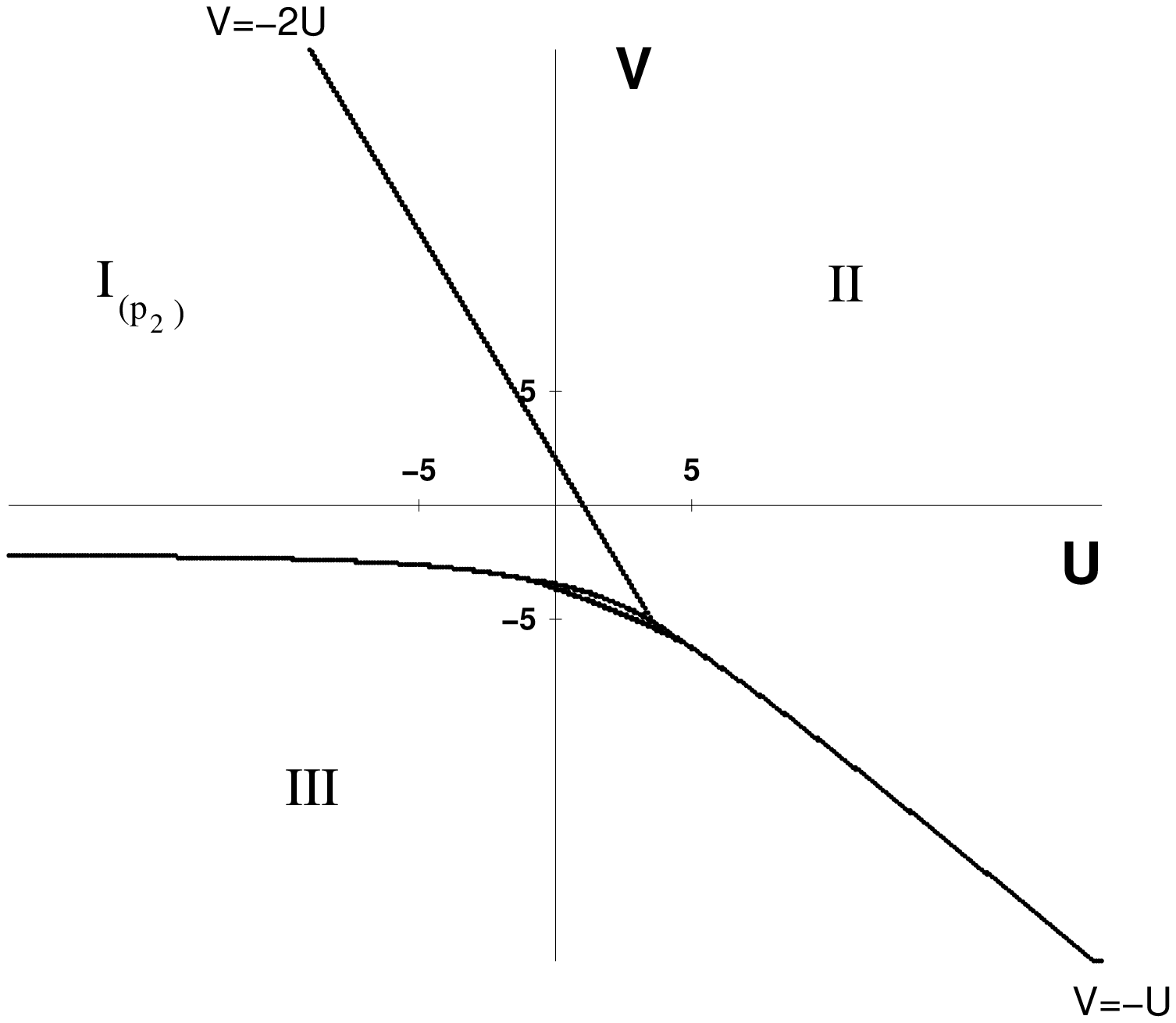} %Paar+4.eps
\caption{Phase diagram for $t_{\perp}=2t_{\parallel}$ including pair hopping 
         for $t_{pair} = 4t_{\parallel}$ with $t_{\parallel}=1$.}
\label{fig8}
\end{figure}

For small negative values of $t_{pair}$ ($|t_{pair}|\le
t_{\parallel}$) the Ising-phase $I$ of the phase diagram in
Fig.~\ref{fig4} with the amplitude $p_2$ splits into two
singlet phases: a symmetric ($p_2$) and an antisymmetric phase ($p_3$)
(see Fig.~\ref{fig7}.a , $t_{pair}=-1$), where the crossover line has
the same gradient ($U=-2V$).  Increasing the amplitude of $t_{pair}$
leads to a pure antisymmetric phase (see Fig.~\ref{fig7}.b ,
$t_{pair}=-4$) in $I$ and also to a strong change of the shape of the
quartet phase.  For small positive values of $t_{pair}$ the general
structure of the phase diagram is preserved (like Fig.~\ref{fig4}).
In the regime of the coupling constants with $t_{pair} \gg
t_{\parallel}$ the quartet phase vanishes (see Fig.~\ref{fig8}).

\begin{figure}
\epsfxsize=0.4\textwidth
(a) \epsfbox{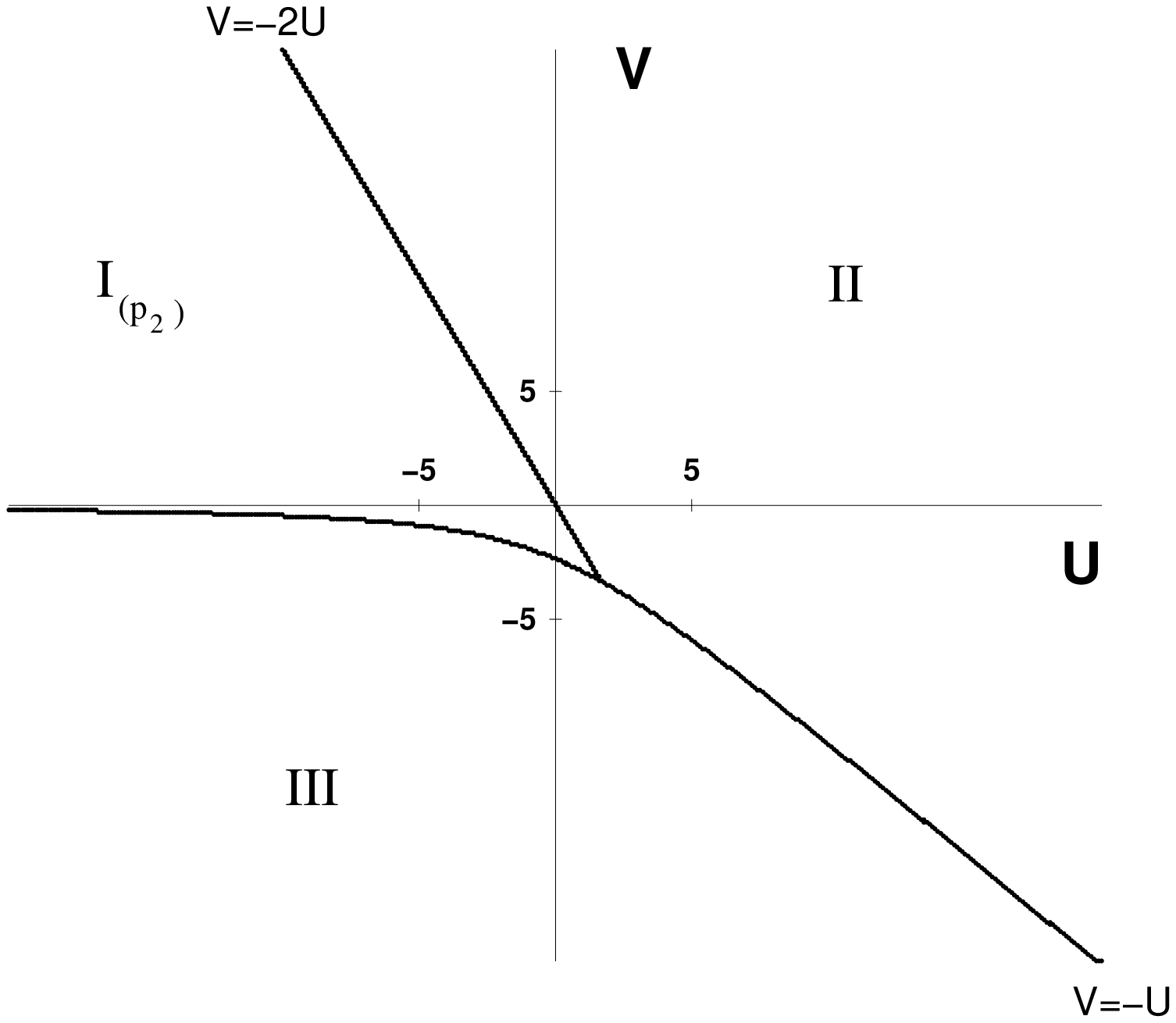} %Quar+4.eps
\hfill
\epsfxsize=0.4\textwidth
(b) \epsfbox{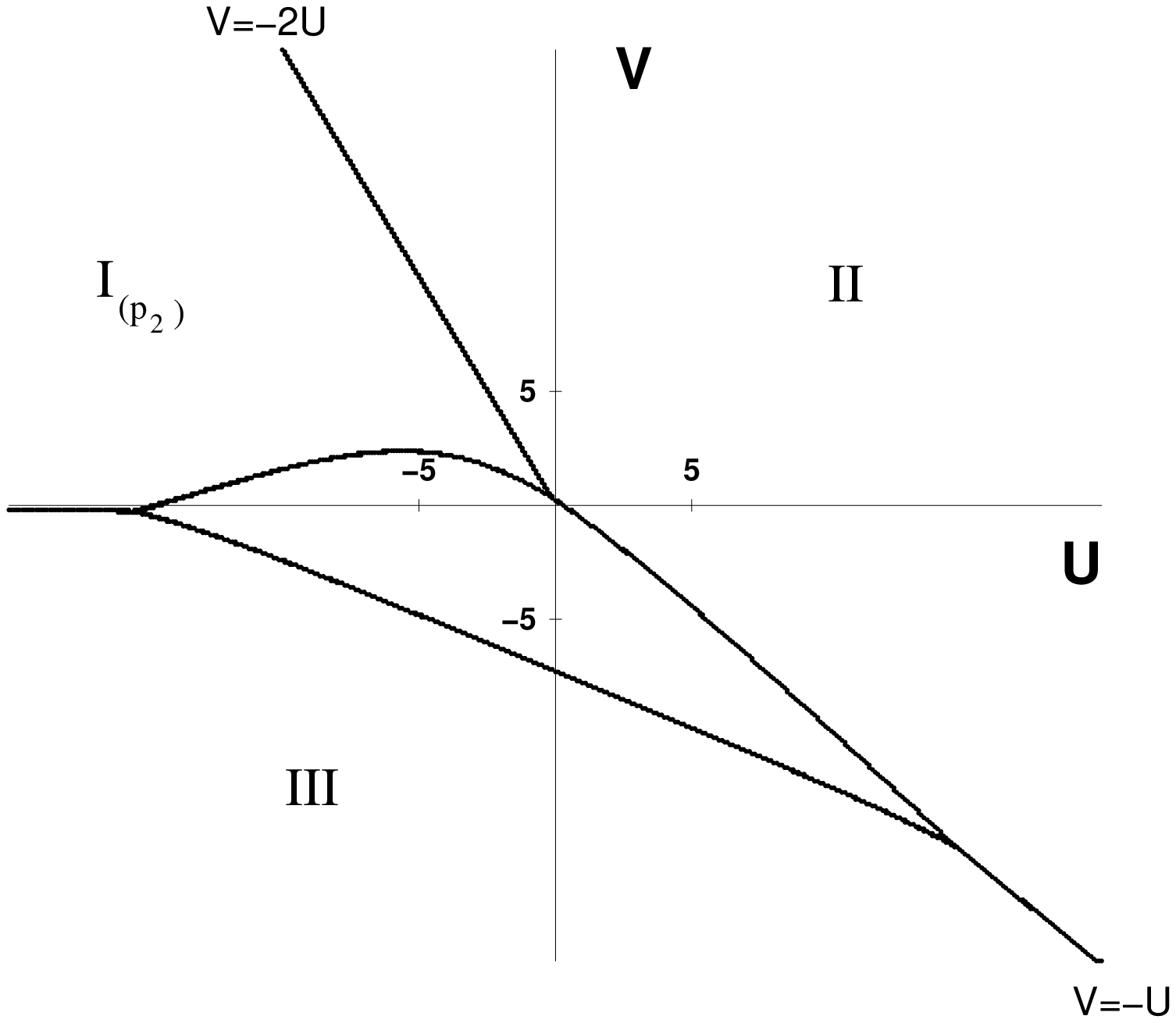} %Quar-4.eps
\caption{Phase diagram for $t_{\perp}=2t_{\parallel}$ including a quartet term 
         with (a) $t_{quar} = +4t_{\parallel}$ and (b) $t_{quar} = -4t_{\parallel}$
         with $t_{\parallel}=1$.}
\label{fig9}
\end{figure}

Other interactions also exhibit strong effects on the phase diagram,
e.g. including a quartet term like (i) which contains a hopping term
on a rung and a bond-charge interaction (see App.~\ref{app:ops}).
\begin{eqnarray}
t_{quar} \left[ (c_{\uparrow}^{\dagger}d_{\uparrow}^{}\!+h.c.)
 (1\!-\!(n_{c\downarrow}\!-n_{d\downarrow})^2)+\uparrow
 \leftrightarrow
 \downarrow\right]
\label{ext2}
\end{eqnarray}
leads to different phase diagrams, depending on the coupling constant.
For positive values of $t_{quar}$ the quartet phase vanishes with
increasing values of the coupling constant (like in Fig.~\ref{fig8})
until there are only the three known phases (see Fig.~\ref{fig9}.a
). For $t_{quar}<0$ the quartet phase grows (see Fig.~\ref{fig9}.b ),
dominated by the symmetric combination ($p_4$) of the states.

The mean deviation in the weak coupling limit in these two special
cases (\ref{ext1},\ref{ext2}) is small compared to the energy
(calculated on a circle with radius $R=0.1$ around $U=V=0$) except for
the value $t_{quar}\le -1$.  The ansatz also provides very good
results in the strong coupling limit ($R \ge 100$), except for the
crossover lines to the superspin phase where the mean deviation is
very large.  The same problem occurs in the intermediate coupling
regime in the quartet phase where the ansatz is not a good eigenstate
of the system.

We expect that including the other interactions on a rung will lead to
similar changes in the ground state phase diagram.

\subsubsection{Two-rung interactions}

In most cases the $SO(5)$-symmetric interactions between two
neighboring terms have a very complex structure but for some of them
we can give simple expressions (see Appendix \ref{app:ops}).  For them
we can calculate the corresponding variational energy, e.g. the
two-pair hopping term leading to an $SO(5)$ singlet-singlet transition
\begin{eqnarray}
t_{2-pair} \left[ d^{\dagger}_{\uparrow}(x) d^{\dagger}_{\uparrow}(y) 
 d^{\dagger}_{\downarrow}(x) d^{\dagger}_{\downarrow}(y)
 c^{}_{\uparrow}(x) c^{}_{\uparrow}(y) 
 c^{}_{\downarrow}(x) c^{}_{\downarrow}(y) +h.c \right]
\end{eqnarray} 
giving $E_{two-pair}\sim 2 t_{2-pair} \left( p_2+p_3\right)^2
\left(p_3-p_2 \right)^2$.  The other $SO(5)$-singlet interactions on
two rungs lead to similar expressions, which will change the phase
diagram (Fig.~\ref{fig4}) in the Ising phase according to the value of
the coupling constant $t_{2-pair}$.

Including an $SO(5)$-quartet interaction in the SZH-model (see
eqn. (\ref{sing-quar})) gives the variational energy
\begin{eqnarray}
 E=t_{qxy}\frac{w-h_1}{2wh_2(w+h_1)} \left(p_2^2-p_3^2\right) 
 \left( p_4^2-p_5^2\right). 
\end{eqnarray}
The phase diagrams obtained for different values of the coupling
constant $t_{qxy}$ are very similar to the phase diagram in
Fig.~\ref{fig4}.  The additional interaction has no significant effect
except for minor changes of the crossover lines.

%%%%%%%%%%%%%%%%%%%%%%%%%%%%%%%%%%%%%%%%%%%%%%%%%%%%%%%%%%%%%%%%%%%%%%
%
\section{Summary and Outlook}
\label{sec:sum}
We have constructed a large class of electronic ladder models with
$SO(5)$ symmetry having finitely correlated ground states and,
consequently, correlation functions exhibiting exponential decay.
These matrix product states have been used to perform a variational
study of the ground state phase diagram of the SZH model \cite{SZH98}
for $t_{\perp}\ge2 t_{\parallel}$.  For vanishing coupling the ground
state of the band insulator is found to be in the class of variational
states and at strong coupling the phases identified by SZH are
reproduced.  In the intermediate coupling regime signatures of a new
phase dominated by local $SO(5)$-quartets are found, and at weak
coupling the $SO(5)$ superspin phase is absent.  Within our approach
it is possible to compute various correlations giving further insights
into the nature of the phases which have been identified.  Finally, we
have introduced various $SO(5)$ symmetric extensions to the SZH model
and discussed their impact on the phase diagram.  In the future we
will include dimerization in the matrix product ansatz for further
studies of the Ising transition\cite{SZH98} in phase I and the
possibility of spontaneous breaking of translational invariance in
exactly solvable models.

%%%%%%%%%%%%%%%%%%%%%%%%%%%%%%%%%%%%%%%%%%%%%%%%%%%%%%%%%%%%%%%%%%%%%%%%%%%%%%%%%%%%%%%%%%

\section*{Acknowledgements}
This work has been supported in parts by the Deutsche
For\-schungs\-gemeinschaft under Grant No.\ Fr~737/3--1.

%%%%%%%%%%%%%%%%%%%%%%%%%%%%%%%%%%%%%%%%%%%%%%%%%%%%%%%%%%%%%%%%%%%%%%
\appendix
\section{The Gamma matrices and the variational wave function}
\label{app:gam}
For the construction of the $SO(5)$-invariant quantities, we have used
the representation of the matrices in Ref. \onlinecite{SZH98}. The
five Dirac $\Gamma$-matrices have the following form
\begin{eqnarray}
\Gamma^1=\left( \begin{array}{cc}  
                 \textstyle{0}  & \textstyle{-i \sigma_y} \\ \textstyle{i \sigma_y}  
&\textstyle{0} 
                \end{array} \right), 
\Gamma^{2,3,4}=\left( \begin{array}{cc}  
                 \textstyle{\vec{\sigma}}  & \textstyle{0} \\ \textstyle{0}  
& \textstyle{\vec{\sigma}^t} 
                \end{array} \right), 
\Gamma^5=\left( \begin{array}{cc}  
                 \textstyle{0}  & \textstyle{\sigma_y} \\ \textstyle{\sigma_y}  
&\textstyle{0} 
                \end{array} \right), 
\end{eqnarray}
where $\vec{\sigma}$ are the Pauli matrices.  The matrices $\Gamma^{ab}$ 
are defined by $\Gamma^{ab}\equiv-\frac{i}{2} [ \Gamma^a,\Gamma^b ] $ and the 
matrix $R$ is given by
\begin{eqnarray}
 R\equiv \left( \begin{array}{rr}  
                 0  & \openone \\ -\openone  & 0 
                \end{array} \right). 
\label{Rmat}
\end{eqnarray}
Using these definitions, it is simple to construct the matrix $g_x$ of
the variational wave function (\ref{ansatz}). It has the following
structure
\begin{equation}
g_x = \left( \begin{array}{c|c}
 \left( \hspace{-2mm}
    \begin{array}{c} \\ \\
       p_6 \Gamma^a n_a  | \Omega \rangle 
     + \sum _{i=1}^3 p_{i}
	 | \widetilde\Psi^{(i)}_{0,0}\rangle \openone \\ \\ \\ 
    \end{array} 
        \hspace{-2mm}
 \right) &
   \begin{array}{c}  
     |q_1\rangle \\ |q_2\rangle \\ |q_3\rangle \\ |q_4\rangle
   \end{array} 
        \hspace{-2mm}
 \\
\noalign{\hrule}
   -|q_3\rangle \quad -|q_4\rangle \qquad |q_1\rangle \qquad |q_2\rangle
 &
 \displaystyle{0}  
       \end{array}
\right)
\label{eq:gx}
\end{equation}
The three $SO(5)$-singlets are included in this ansatz only on the
main diagonal elements.
$|\widetilde\Psi^{(1)}_{0,0}\rangle\equiv|\Psi^{(1)}_{0,0}\rangle$
from (\ref{sing1}) and $|\widetilde\Psi^{(2,3)}_{0,0}\rangle$ are the
symmetric and an antisymmetric combinations of the two other singlets
\begin{equation}
  |\widetilde\Psi^{(2,3)}_{0,0}\rangle
  \sim \left( \Psi^{\dagger}_{\alpha} R_{\alpha \beta} \Psi^{\dagger}_{\beta} \mp 
   \Psi^{}_{\alpha} R_{\alpha \beta} \Psi^{}_{\beta}  \right) 
  |\Omega \rangle. 
\end{equation}

The quartets enter the matrix $g_x$ in $|q_\alpha\rangle = p_4
|Q_\alpha^+\rangle + p_5 |Q_\alpha^-\rangle$ where
$|Q^\pm_\alpha\rangle$ are the symmetric and antisymmetric
combinations of (\ref{quar1})
\begin{equation}
\left| Q_{\alpha}^{\pm} \right\rangle \sim
\left\{
    \left| { \downarrow \atop - } \right\rangle 
\pm \left| { - \atop \downarrow } \right\rangle , 
    \left| { \uparrow \atop - } \right\rangle 
\pm \left| { - \atop \uparrow } \right\rangle , 
    \left| { \uparrow \atop \uparrow \downarrow } \right\rangle 
\pm \left| { \uparrow \downarrow \atop \uparrow } \right\rangle , 
    \left| { \downarrow \atop \uparrow \downarrow } \right\rangle 
\pm \left| { \uparrow \downarrow \atop \downarrow } \right\rangle 
\right\}\ .
\label{eq:quar}
\end{equation}
They are arranged in the right column and the lowest row of
(\ref{eq:gx}) in such a way that in the product $g_x g_{x+1}$ one has
$SO(5)$-singlets on the diagonal only.

%%%%%%%%%%%%%%%%%%%%%%%%%%%%%%%%%%%%%%%%%%%%%%%%%%%%%%%%%%%%%%%%%%%%%%
\section{SO(5)-symmetric operators on one resp. two rungs}
\label{app:ops}
We present here a selection of various $SO(5)$-symmetric terms on a
single and on two rungs
(\ref{app:ops:si},\ref{app:ops:two}). Furthermore, a list of terms is
given for which our matrix product ansatz (\ref{ansatz}) would be the
lowest energy state (\ref{app:ops:gs}).
\subsection{Single rung interactions}
\label{app:ops:si}
We now present all possible $SO(5)$-symmetric terms on a rung.  Their
general construction is done in terms of projection operators on the
different $SO(5)$-multiplets.  Expressed through electronic operators
most of them are already known from the SZH-model (\ref{SZH-ham}) and
an additional biquadratic exchange.  As a shorthand notation we
introduce
\begin{eqnarray} 
  [U,V,J,\alpha] &\equiv&
  \; U \; \left( ( n_{c\uparrow}(x)-\frac{1}{2} )
       ( n_{c\downarrow}(x)-\frac{1}{2} ) + ( c \rightarrow d )\right)
\nonumber\\
 && \quad
      + \;V \; (n_c(x)-1)(n_d(x)-1) 
      +\; J \; \vec{S_c}(x) \vec{S_d}(x)  
\\
 &&\quad     + \alpha (\vec{S_c}(x)\vec{S_d}(x))^2
\nonumber
\end{eqnarray}
In addition we find various single electron and pair hopping terms
together with bond-charge type interactions.  Using the notation of
Sect.~\ref{sec:states} for the projection operators on a rung we
obtain the following terms by projection on the singlets
\begin{eqnarray}
 \hat{P}_{0,0}^{1,1}&=& | \Psi_{0,0}^{(1)}\rangle \langle\Psi_{0,0}^{(1)}|
= \left[0,0,-\frac{1}{3},\frac{4}{3}\right] \\
 \hat{P}_{0,0}^{2,2}&=&
 \left[\frac{1}{2},-\frac{1}{4},\frac{2}{3},\frac{4}{3}\right ]+\frac{1}{2} 
 \left[ n_{d\uparrow}n_{d\downarrow} (1-n_c) - c \leftrightarrow d
\right]   
\\
\hat{P}_{0,0}^{3,3}&=&\left[\frac{1}{2},-\frac{1}{4},\frac{2}{3},\frac{4}{3}
 \right]-\frac{1}{2} 
 \left[ n_{d\uparrow}n_{d\downarrow} (1-n_c) - c
\leftrightarrow d \right] 
 \\
 \hat{P}_{0,0}^{1,2}&+&\hat{P}_{0,0}^{2,1}=\frac{1}{\sqrt{2}} \left[ (c_{\uparrow}^{\dagger}d_{\uparrow}^{}+h.c.) n_{d\downarrow} (n_{c\downarrow}-1) + \uparrow \leftrightarrow \downarrow \right] \\
 \hat{P}_{0,0}^{1,3}&+&\hat{P}_{0,0}^{3,1}=-\frac{1}{\sqrt{2}} \left[ (c_{\uparrow}^{\dagger}d_{\uparrow}^{}+h.c.) n_{c\downarrow} (n_{d\downarrow}-1) + \uparrow \leftrightarrow \downarrow \right]  \\
 \hat{P}_{0,0}^{2,3}&+&\hat{P}_{0,0}^{3,2}=
d_{\uparrow}^{\dagger}d_{\downarrow}^{\dagger}c_{\uparrow}c_{\downarrow}
+h.c.  \ .
\end{eqnarray}
The projection operators on the quartet states read
\begin{eqnarray}
 \sum_{\mu=1}^4 \hat{P}_{4,\mu}^{1,1}&=& \left[0,0,-\frac{8}{3},-\frac{16}{3}\right] 
     +(1-n_{c\uparrow}n_{c\downarrow})n_d +n_{d\uparrow}n_{d\downarrow} (n_c-2) \\
 \sum_{\mu=1}^4 \hat{P}_{4,\mu}^{2,2}&=& \left[0,0,-\frac{8}{3},-\frac{16}{3}\right] 
     +(1-n_{d\uparrow}n_{d\downarrow})n_c +n_{c\uparrow}n_{c\downarrow} (n_d-2)   \\
 \sum_{\mu=1}^4 \hat{P}_{4,\mu}^{1,2}&+&\hat{P}_{4,\mu}^{2,1}=
     \left[ (c_{\uparrow}^{\dagger}d_{\uparrow}^{}\!+h.c.)
 (1\!-\!(n_{c\downarrow}\!-n_{d\downarrow})^2)+\uparrow \leftrightarrow \downarrow\right],
\end{eqnarray}
and finally, projection on the quintet gives
\begin{eqnarray}
 \sum_{\mu=1}^5 \hat{P}_{5,\mu}=\left[
 1,\frac{1}{2},\frac{13}{3},\frac{20}{3} \right] \ .
\end{eqnarray}

\subsection{Interactions between neighboring rungs}
\label{app:ops:two}
Equivalently, the $SO(5)$-symmetric expressions on two rungs can be
classified. The choice of the basis on the two-rung system is very
important for the structure of the $SO(5)$-symmetric terms. Using the
simplest combination the product of an $SO(5)$-singlet on one rung and
another $SO(5)$ multiplet on the other gives for a projection operator
e.g.
\begin{eqnarray}
 \sum_{\mu=1}^5 \hat{P}_{5,\mu}^{1,1}(x,y)=\hat{P}_{0,0}^{1,1}(x) \sum_{\mu=1}^5 \hat{P}_{5,\mu}(y)
\end{eqnarray}
for the product of an $SO(5)$-singlet on rung $x$ and an
$SO(5)$-quintet on $y$, where $\hat{P}_{d,\mu}^{k,l}$ is defined in
section \ref{sec:states}.  The numbers $k$ and $l$ in
$\hat{P}_{d,\mu}^{k,l}$ depend on the way the different multiplets on
the rungs are labelled.  Another example is an operator projecting on
an $SO(5)$-singlet on each of the rungs
\begin{eqnarray}
 \hat{P}_{0,0}^{2,2}(x,y)=|\Psi^{(2)}_{0,0}(x,y) \rangle \langle \Psi^{(2)}_{0,0} (x,y)| = 
\hat{P}_{0,0}^{1,1}(x) \hat{P}_{0,0}^{2,2}(y).
\end{eqnarray}
(see equation (\ref{sing1}) for the definition of the wave functions).
All projection operators of states consisting of at least one
$SO(5)$-singlet on a rung can be decomposed in the same manner.  For
some of these operators a compact representation in terms of electron
operators is possible.  As an example consider the operator
\begin{eqnarray}
 c^{\dagger}_{\uparrow}(x) c^{\dagger}_{\downarrow}(x) 
 d^{\dagger}_{\uparrow}(y) d^{\dagger}_{\downarrow}(y)
 c^{}_{\uparrow}(y) c^{}_{\downarrow}(y) 
 d^{}_{\uparrow}(x) d^{}_{\downarrow}(x) +h.c. 
\label{pairhopp}
\end{eqnarray}
describing pair exchange between two neighboring rungs.  It causes a
transition between two $SO(5)$ singlet states and can be written as
\begin{equation}
\sim \hat{P}_{0,0}^{3,2}(y) \hat{P}_{0,0}^{3,2}(x)+h.c.\ .
\end{equation}
Other $SO(5)$ singlet-singlet transitions of this type are
\begin{eqnarray}
&& d^{\dagger}_{\uparrow}(x) d^{\dagger}_{\uparrow}(y) 
 d^{\dagger}_{\downarrow}(x) d^{\dagger}_{\downarrow}(y)
 c^{}_{\uparrow}(x) c^{}_{\uparrow}(y) 
 c^{}_{\downarrow}(x) c^{}_{\downarrow}(y) +h.c \ ,
\nonumber\\
&& N_d(y)n_{c\uparrow}^{}(y)n_{c\downarrow}^{}(y)
\left[ d^{\dagger}_{\uparrow}(x)d^{\dagger}_{\downarrow}(x)
c^{}_{\uparrow}(x) c^{}_{\downarrow}(x) + h.c \right]\ ,
\\
&& N_c(x)n_{d\uparrow}^{}(x)n_{d\downarrow}^{}(x)
\left[ c^{\dagger}_{\uparrow}(y)c^{\dagger}_{\downarrow}(y)
d^{}_{\uparrow}(y) d^{}_{\downarrow}(y) + h.c \right]\ ,
\nonumber
\end{eqnarray}
where
\begin{equation}
N_{\alpha}(y)= \left( 1-n_{\alpha\uparrow}^{}(y)-n_{\alpha\downarrow}^{}(y)
               +n_{\alpha\uparrow}^{}(y)n_{\alpha\downarrow}^{}(y) \right),
\hspace{1cm} \alpha \in \left\{ c,d \right\}
\label{Nd}
\end{equation}

Similar terms are obtained from projection operators on direct
products of an $SO(5)$-singlet on one and an $SO(5)$-quartet on the
other rung, e.g.
\begin{eqnarray}
&& \left[ \left( \left(n_{c\uparrow}(x)-n_{d\uparrow}(x)\right)^2 -1 \right)  
         c_{\uparrow}^{\dagger}(y)c_{\downarrow}^{\dagger}(x)c_{\downarrow}^{\dagger}(y)
         d_{\uparrow}^{}(y)d_{\downarrow}^{}(x)d_{\downarrow}^{}(y) 
   \right. 
\label{sing-quar} \\ \nonumber
+ && \left.\left( \left(n_{c\downarrow}(x)-n_{d\downarrow}(x)\right)^2 -1 \right) 
         c_{\uparrow}^{\dagger}(x)c_{\uparrow}^{\dagger}(y)c_{\downarrow}^{\dagger}(y)
         d_{\uparrow}^{}(x)d_{\uparrow}^{}(y)d_{\downarrow}^{}(y) 
    \right] +h.c
\end{eqnarray}

The projection operators on the remaining $169$ states with a
structure similar to (\ref{quar2}) cannot easily be decomposed in this
way.  They are significantly more complex, generically their expansion
into electronic operators produces complicated bond-charge interaction
terms.  Still, forming suitable linear combinations of such terms can
lead to simpler $SO(5)$-symmetric terms on two rungs, e.g.  the pair
hopping term in (\ref{H-tpar}) or a diagonal hopping term

\begin{equation}
\sum_{\left< x,y \right>}  
\left[  d_{\sigma}^{\dagger}(x)c_{\sigma}(y) 
      - c_{\sigma}^{\dagger}(x)d_{\sigma}(y) + h.c. \right].
\end{equation}

%%%%%%%%%%%%%%%%
\subsection{SO(5)-symmetric Hamiltonians with exact Ground States}
\label{app:ops:gs}
At the end of section \ref{sec:ext} we claimed that a general Hamiltonian 
where our ansatz (\ref{ansatz}) is the ground state of the system is given 
by
\begin{eqnarray} 
     h_{x,x+1}&=& \sum_{k,l=1}^4 \lambda_{16}^{(k,l)}
	 \sum_{\mu=1}^{16} \hat{P}_{16,\mu}^{k,l}
  + \sum_{\mu=1}^{14} \hspace{1mm} \lambda_{14} \, \hat{P}_{14,\mu}^{} 
  + \mbox{additional terms.}
\label{exactHam}
\end{eqnarray}
The coupling constants have to be chosen such that $\lambda_{14}>0$
and the matrix $\lambda_{16}$ of coupling constants is positive
definite.  This implies that $E=0$ is a lower bound on the spectrum
and therefore the state (\ref{ansatz}) --- having zero energy by
construction --- will be a ground state.
The additional terms in (\ref{exactHam}) are the projection operators
on the remaining multiplets not present in the matrix product wave
function.  
For example the projection operator on one of the $SO(5)$-singlets not
present in this product reads
\begin{eqnarray}
 \lambda_0^{(k,l)}  && \left[-c_{\uparrow}^{\dagger}(x)
   c_{\uparrow}^{\dagger}(y)
   c_{\downarrow}^{\dagger}(x)c_{\downarrow}^{\dagger}(y)
   d_{\uparrow}^{}(x)d_{\uparrow}^{}(y)
   d_{\downarrow}^{}(x)d_{\downarrow}^{}(y) 
 + n_{d\uparrow}(x)n_{d\uparrow}(y)n_{d\downarrow}(x)n_{d\downarrow}(y)
   N_c(x)N_c(y) \right. \\\nonumber
 && + \left. (c \leftrightarrow d) \right]
\end{eqnarray}
Just as $\lambda_{16}$ above the matrices $\lambda_d\; (d=0,4,5,10)$,
coupling the projection operators on the remaining multiplets have to
be chosen to be positive definite.

%%%%%%%%%%%%%%%%%%%%%%%%%%%%%%%%%%%%%%%%%%%%%%%%%%%%%%%%%%%%%%%%%%%%%%
%
\section{Correlation functions}
\label{app:cor}
The calculation of expectation values between matrix product states is
straightforward using a transfer matrix method (see e.g. Ref. 
\onlinecite{ksz:92}):

To this end we define a $25 \times 25$ transfer matrix $G$ on a rung
\begin{eqnarray}
 G_{\alpha_1,\alpha_2} \sim G_{(i_1,j_1),(i_2,j_2)} 
 \equiv g^{\dagger}_{(i_1,i_2)} g^{}_{(j_1,j_2)}
\end{eqnarray} 
with the indices
\begin{eqnarray*}
  \alpha_1=1,\ldots,25 \leftrightarrow (11),\ldots,(15),(21),\ldots,(55).
\end{eqnarray*}
In terms of $G$ the norm of the ground state can be written as
\begin{eqnarray}
 \langle \Psi_0 \mid \Psi_0 \rangle = \mbox{Tr } G^L = \sum_{i=1}^{25} \lambda^L_{i}
\end{eqnarray}
where $\lambda_i$ are the eigenvalues of $G$. In the thermodynamic
limit ($L \to \infty$) the largest eigenvalue $\lambda_1$ dominates
this expression and we obtain $\langle \Psi_0 | \Psi_0 \rangle \sim
\lambda_1^L$.  Similarly, one-point correlators of an operator $O$ are
\begin{eqnarray}
 \langle O \rangle = \frac{1}{\lambda_{1}} \langle e_{1} | Z(O) | e_{1} \rangle
\end{eqnarray} 
and a two-point correlation function reads
\begin{eqnarray}
 \langle O_1^{\dagger} O_r \rangle\! =\! \sum_{n=1}^{25} \frac{1}{\lambda_n^2} \left( 
  \frac{\lambda_n}{\lambda_{1}} \right)^r \! \langle  e_{1} | Z(O_1)  | e_{n} \rangle 
 \langle e_{n} | Z(O_r) | e_{1} \rangle.  
\end{eqnarray} 
Here $| e_{n} \rangle$ are the eigenvectors with eigenvalue
$\lambda_n$ of $G$ and $Z(O_i)\sim g^{\dagger}O_ig$ is the transfer
matrix related to the operator $O_i$.  With the matrix (\ref{eq:gx})
the largest eigenvalue of $G$ is given by
\begin{eqnarray}
 \lambda_1=\frac{1}{2} \left( h_1+w \right)
\end{eqnarray}
with $h_1=5p_6^2+p_1^2+p^2_2+p_3^2$, $h_2=p^2_4+p^2_5$ and
$w=\sqrt{h_1^2+16h_2^2}$.

This enables us to calculate the expectation values of any operator
acting on a single or two rungs, respectively.  For example, we find
\begin{eqnarray}
 \langle \vec{S} \rangle=0, \hspace{1cm} \left\langle (S^i)^2 \right\rangle=
 \frac{w-h_1+8p_6^2}{4w}  
\end{eqnarray}
for local magnetic moments and
\begin{eqnarray} 
 \langle c_{g(u)}^{\dagger}(x) c_{g(u)}^{}(x)  \rangle &=&\frac{1}{w} \left[ \frac{w-h_1}{h_2} 
(p_5^2+3p_4^2) + 2h_1 \mp 4p_1p_2) \right]  \\
 \langle c_{\alpha}^{\dagger}(x) c_{\beta}^{}(x)  \rangle &=&\frac{1}{w} \left[ \frac{w-h_1}{h_2} 
(2p_4p_5)-4p_2p_3 \right], \hspace{1cm} \alpha\ne\beta 
\end{eqnarray}
for electronic expectation values.  Here, $\alpha,\beta \in \left\{ g,u
\right\}$ and $ c_{g,u}(x)=(c_{\uparrow}(x)+c_{\downarrow}(x)) \pm
(d_{\uparrow}(x) + d_{\downarrow}(x))$.

Correlations between the total spin on two rungs decay exponentially
\begin{eqnarray}
 \langle \vec{S_1} \vec{S_r} \rangle=-\frac{3}{4w(h_1+w)} \left(\frac{h_1-4p_6^2}{\lambda_{1}}
 \right)^r \left(\frac{w-h_1+8p_6^2}{h_1-4p_6^2} \right)^2
\end{eqnarray}
as expected for finitely correlated states.  Spin-spin correlations
between individual sites on rungs separated by a distance $r$ can be
expressed as
\begin{eqnarray} 
\langle \vec{S}_{\alpha}(r)\vec{S}_{\beta}(0) \rangle 
  = A_{\alpha\beta}(\{p_i\}) \left( 
   \frac{h_1-4p_6^2}{\lambda_1}\right)^r  
  + B_{\alpha\beta}(\{p_i\}) \left( \frac{h_1-8p_6^2}{\lambda_1}
                     \right)^r, 
\end{eqnarray}
where the amplitudes $A_{\alpha\beta}(\{p_i\})$ and
$B_{\alpha\beta}(\{p_i\})$ depend on the choice of $\alpha$ and
$\beta$, i.e.\ whether correlators of spins on the same or on
different legs of the ladder are considered.
Analogously, one can study electronic correlations, e.g.\
\begin{eqnarray} 
&& \langle c_g^{\dagger}(r) c_u^{\dagger}(r) c_g^{}(0) c_u^{}(0) \rangle =
 -\frac{8}{3} \langle \vec{S_1} \vec{S_r} \rangle
\nonumber\\
&& \langle c_{g,u}^{\dagger}(r) c_{g,u}^{}(0) \rangle = 
 C_{g,u}(\{p_i\}) \left(\frac{h_2}{\lambda_1}\right)^r + D_{g,u}(\{p_i\}) 
\left(\!-\frac{h_2}{\lambda_1}\right)^r\!. 
\end{eqnarray}
with the amplitudes $C_{g,u}$ and $D_{g,u}$. 

%\setlength{\baselineskip}{13pt}
%\bibliographystyle{prsty}
%\bibliography{Base} %base,heisenberg,peierls,Base}

%%%%%%%%%%%%%%%%%%%%%%%%%%%%%%%%%%%%%%%%%%%%%%%%%%%%%%%%%%%%%%%%%%%%%%
%%%%%%%%%%%%%%%%%%%%%%%%%%%%%%%%%%%%%%%%%%%%%%%%%%%%%%%%%%%%%%%%%%%%%%
%%%%%%%%%%%%%%%%%%%%%%%%%%%%%%%%%%%%%%%%%%%%%%%%%%%%%%%%%%%%%%%%%%%%%%

%\end{multicols}

\end{document}